# STUDENT SENSEMAKING ON ELECTROSTATICS PROBLEMS INVOLVING THE METHOD OF IMAGES THROUGH THE LENS OF EPISTEMIC GAME FRAMEWORK

Jaya Shivangani Kashyap and Chandralekha Singh

*Department of Physics and Astronomy, University of Pittsburgh, 3941 O'Hara Street, Pittsburgh, Pennsylvania, 15260, USA*

Understanding the mechanisms of student sensemaking while navigating the physics problem-solving process can play an important role in developing approaches to helping students become proficient problem solvers. Since most prior studies on student sensemaking in physics have focused on introductory physics, this study sheds light on a relatively unexplored research area involving advanced physics students' sensemaking. We conducted individual interviews with graduate students to investigate their sensemaking while solving upper-level undergraduate electrostatics problems in the context of the method of images. Analysis through the lens of the epistemic game framework proposed by Tuminaro and Redish shows that the central part of student sensemaking can be mapped on to the Pictorial Analysis game with other epistemic games also playing a role, but the ontological components of some games require refinement. The study involves an in-depth comparison of sensemaking on related but distinct problems for individual graduate students and comparison between multiple students solving the same problems. It provides insights into how graduate students transfer and translate strategies between different problems while sensemaking. Graduate students showed persistence in activating different knowledge resources and consistent use of reasoning primitives across different problems such as a "balancing" primitive multiple times while sensemaking. The study investigated the impact of small nudging and scaffolding that resulted in significant improvements in graduate students' problem-solving approaches. Another novel finding is that graduate students' problem-solving approaches significantly improved with successive attempts and multiple drawings for the same problem and across problems. Often, graduate students initially drew more three-dimensional figures with conducting planes given in the problem statement to visualize and understand the problem; they also drew diagrams from different viewing angles which helped in improving their physical understanding and comprehension of the symmetry of the underlying problem.

# I. INTRODUCTION, FRAMEWORK, AND GOAL

Learning electricity and magnetism [1] is challenging for students. Solving physics problems at all levels requires students to have a good understanding of relevant mathematics and the ability to integrate these mathematical concepts [2-20] seamlessly with the physics concepts. However, even after being equipped with sophisticated mathematical tools and physics concepts, many students struggle to integrate physics and mathematics appropriately during problem solving [21-28]. These issues can be better understood by observing student sensemaking as it can help understand how students use their knowledge, what resources are activated in different situations, and what could be activated if a certain level of guidance and scaffolding support is provided. Here, we build on a prior study on physics graduate students' sensemaking in the context of Laplace's equation [29] and discuss graduate student sensemaking during interviews conducted as a part of the development and validation of a tutorial on the method of images (MoI) for the upper–level undergraduate electricity and magnetism [30-33] course. Since most prior studies on sensemaking have focused on introductory physics, this novel study sheds light on a relatively unexplored research area involving advanced physics students' (graduate students') sensemaking.

In electricity and magnetism, the method of images (MoI) is a powerful technique used to solve certain kinds of boundary value problems. For example, for a problem consisting of a large thin grounded and a point charge present near it (as shown in Fig. 1(a)), the calculation of the potential in the region where the original charge is present (shown as the shaded region in Fig. 1(a)) can be complex. The presence of the point charge above the grounded conducting plane induces a charge distribution on the conductor's surface. Thus, calculating the potential at a point in space, e.g., using the direct integration method could be challenging. The corollary of the first uniqueness theorem states [34] that the potential ($V$) in a volume $\mathcal{V}$ is uniquely determined if the charge density throughout the region and the value of $V$ on all the boundaries are known. Taking advantage of this corollary, we can replace our original complicated problem with the new configuration as shown in Fig. 1(b). The grounded conductor can be replaced by an image charge in the excluded region (the region below the conductor, where the potential is not of interest), yielding the same charge density and the boundary conditions as the original configuration in the desired region, where the potential is to be calculated. Using this approach, the complicated problem can be replaced by a simplified problem, which could be solved by a student who knows the mathematical expression for the potential due to a point charge and the superposition principle. The MoI is extremely helpful to solve such boundary value problems including conductors and charge distributions, so it is important for students to learn this approach. Students usually learn this method in the upper-level electricity and magnetism course, but they do not always use it appropriately while solving problems. To provide additional aid to the students, we developed and validated a tutorial along with a pretest and posttest on the MoI. As a part of this process, we collected qualitative data by conducting interviews with the physics graduate (grad) students. We did not interview the undergraduate students who were enrolled in the upper-level electricity and magnetism course at that time (when the grad students were interviewed), because they had not had instruction in relevant concepts. We observed that graduate students showed sensemaking patterns that reflected that they are still developing expertise. In particular, these patterns are consistent with what one would expect from upper-level undergraduates in similar contexts, who have learned these concepts in an upper-level electricity and magnetism course. Thus, graduate students' sensemaking patterns are unlikely to be significantly different from those of upper-level undergraduates for the purposes of sensemaking in this study. Therefore, although the advanced students who participated in this study are physics graduate students, the findings are likely to be relevant for advanced students, who are upper-level undergraduates and have taken an electricity and magnetism course beyond introductory physics.

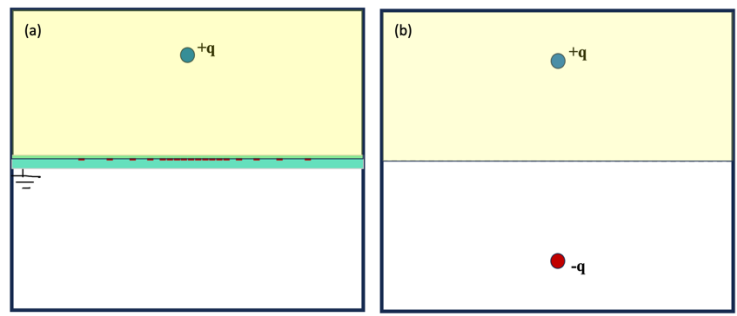

Fig. 1. (a) Charge present near a grounded conducting plane (left), (b) conductor is replaced by an image charge (right).

Sensemaking [35-43] is the process of figuring out how to solve a problem. In the context of physics, when students are provided with a novel problem that they might not have solved before and are asked to solve it within a given timeframe, they try to make sense while navigating through the problem-solving process [44-51]. Sensemaking can help in creating connections within students' pre-existing knowledge and generating new knowledge. Students might activate and access physics and mathematics concepts learned in formal contexts as well as their intuitive knowledge from everyday life that they already know as resources to create connections between what they know and what the problem is asking for to reduce the gap between these in the context of problem-solving [52-65]. Students' epistemological beliefs about physics (e.g., whether it is a collection of facts and formulas, whether the instructor is an authority) may play a role in what knowledge is activated and how it is accessed during problem-solving [66]. The process of sensemaking in which students use their knowledge resources to solve a problem can lead to generating new knowledge that students did not know before (regardless of whether they are consistent with physical laws and concepts) [67].

Observation of student sensemaking can help understand how students think while solving a physics problem. It can help understand what knowledge resources get activated in a particular situation and what additional resources could potentially be activated if students are provided with a certain level of guidance and scaffolding during the problem-solving process. We note that diSessa [36, 45] introduced the notion of phenomenological primitives or p-prims that arise from the activation and interaction of multiple cognitive resources that involve the blending of intuitive elements from everyday life experiences. He and other researchers [36, 44, 45, 68] emphasized that since students' knowledge is in pieces (fragmented), their sensemaking will only show local coherence and lack global coherence. For example, a student can activate the knowledge resource pertaining to the closer is stronger p-prim to argue that the earth is closer to the sun in summer when it is warmer. In the context of introductory physics, they emphasized that students' sensemaking during physics problem-solving can be enhanced if they are provided support in learning to integrate their formal and intuitive knowledge resources [36, 44, 45, 68]. Tuminaro and Redish expanded the idea of p-prims to reasoning primitives, which are 'group cognitive structures at their different levels of abstraction' [66] and do not need to be derived from everyday life experiences. For example, 'agent causes effect' is an abstract reasoning primitive that can be mapped into 'force as mover' or 'force as spinner' p-prims [66].

Similar to our study on advanced students' sensemaking in the context of Laplace's equation [29], we investigated the sensemaking of physics graduate students through the lens of the epistemic game framework adapted initially by Tuminaro and Redish [66] in the physics context from the framework originally proposed by Collins and Ferguson [69]. In the context of physics, problem solving involves playing different epistemic games. Depending upon the situation and the context of the problem [70], it can involve different types of games being played by the students. Tuminaro and Redish defined an epistemic game as 'a coherent activity that uses particular kinds of knowledge and processes associated with that knowledge to create knowledge or solve a problem' [66]. The structural components of epistemic games are entry conditions, moves, and ending conditions [66]. The entry and ending conditions are defined as the conditions that lead to the beginning and the ending of a particular game, respectively. For a given problem, the prompt of the question makes students enter a particular kind of epistemic game. The ending conditions could be figuring out a mathematical expression, obtaining a numerical answer, providing a reasoning for the problem, or giving up on the problem-solving process because the student could not figure out how to proceed. Students use different methods or strategies, which could be defined as moves, to reach the ending conditions [66]. Different games can have different moves depending on the requirements of the problem.

Tuminaro and Redish do not claim that the epistemic games or all possible moves they mentioned within each game for introductory physics problem-solving span a complete list of games useful in other contexts, but their set of games has a relatively broad applicability [66]. Although their dataset focuses on introductory-level students, we focus here on graduate students solving upper-level problems that can be solved efficiently using the MoI. We find that in some cases, the ontological components of the epistemic games proposed by Tuminaro and Redish need to be interpreted somewhat differently in the context of the upper-level students' problem solving than those for introductory-level students' problem solving. The broader structure of the game remains the same for both introductory level and upper-level students' problem solving. For example, in many of Tuminaro and Redish's epistemic games, students only activated intuitive knowledge resources that were derived exclusively from everyday life experiences (no connection with formal learning). However, in the sensemaking of advanced students in the context of MoI using epistemic games, we find that intuitive knowledge resources that students activated during problem-solving can have components learned in formal learning contexts. With these types of tweaks to the ontological components of the epistemic games, here we describe the findings from the mapping of some of the sensemaking episodes from our interviews with physics graduate students onto the epistemic game framework.

Tuminaro and Redish classified the epistemic games into six categories: Mapping Mathematics to Meaning, Mapping Meaning to Mathematics, Physical Mechanism, Pictorial Analysis, Transliteration to Mathematics and Recursive Plug and Chug [66]. We found that for the MoI, the Pictorial Analysis epistemic game is the most prominent game played by students as expected because the approach involves a deep analysis involving diagrams. This game includes the determination of the target concept, selection of an external representation (picture), a narration of a conceptual story about the physical situation based on the spatial relation among the objects and finally filling up the "slots" (components) in the representation (picture/diagram).

However, we also find that during interviews, advanced students played other games within the moves of the Pictorial Analysis game, or they switched between different games while solving a problem. The other games we observed in the investigation described here are Mapping Meaning to Mathematics, Mapping Mathematics to Meaning, Transliteration to Mathematics and Physical Mechanism.

As Tuminaro and Redish defined, Mapping Meaning to Mathematics game includes the development of a story about the physical situation in which students rely on their own conceptual understanding rather than the fundamental physics [66]. The quantities are then translated into mathematical entities andrelated back to the physical story. At the end, symbolic manipulation is carried out, and the story is eventually evaluated. We adapted this game onto our dataset from graduate student interviews involving the MoI and found that these students often activated knowledge resources that were valid physics concepts in these problems while sensemaking, but they were not relevant in the context used. As discussed by Tuminaro and Redish, introductory physics students did not rely on the fundamental physics concepts while playing this game. Our generalization of the game includes cases in which students could be using the correct physics concept but in an incorrect context while playing this epistemic game. For example, we discuss the Mapping Meaning to Mathematics game in cases in which the interviewed graduate students played this game to obtain the expression for the potential and used boundary conditions to verify the correctness of their approach but used incorrect image charges or incorrect distances. The Mapping Mathematics to Meaning game includes the identification of the target concept. It begins with a physics equation, and the conceptual story is developed based on that. The structure of the Transliteration to Mathematics game is slightly different, as the solution of a novel problem is obtained by mapping the quantities of the current problem with the solution pattern of another problem that was solved by the student before. Tuminaro and Redish described that introductory students who play the Transliteration to Mathematics epistemic game do not have a conceptual understanding of the worked example but use it from memory to generate solutions to new problems that resemble problem solutions they have seen in the past. In the episodes discussed here of graduate students' sensemaking, we extend the context of the Transliteration to Mathematics game. In this game, we include cases in which students may have a reasonable conceptual understanding of the solution they have encountered in the past and use it as a reference to find the solution to another novel problem, but this still might not give the correct form of the solution due to some errors. The Physical Mechanism game involves the construction of a physically coherent and descriptive story based on the intuitive sense of the physical mechanism. Tuminaro and Redish described that in this game, students draw on their intuitive knowledge base rather than formal knowledge. They described that this intuitive knowledge comes from everyday experiences. As we discuss sensemaking in the context of the MoI problems in upper-level electricity and magnetismcourses, it is difficult for students to connect integration, differentiation, image charges, etc. to everyday life experiences in most cases. We consider that for advanced students in these contexts; their intuitive knowledge can come from patterns identified from past learning. For example, the physics or math concepts that they have learned in other physics and mathematics classes or earlier introductory courses become part of their intuitive physics problem solving toolbox. Interestingly, we find students to be working mindfully during these interview sessions, such that there is always some local coherence visible during their sensemaking. The episodes discussed here do not show any evidence of students playing the Recursive Plug and Chug game, which was observed in Tuminaro and Redish's introductory students' sensemaking. We note that some parts of the episodes can also be mapped onto other epistemic games not discussed here. In particular, here we primarily focus on big picture games, which are prominent while sensemaking.

With this background and framework, the goal of this research is to shed light on physics grad students' sensemaking through the epistemic game framework of Tuminaro and Redish, with modifications in the ontological components of some of the games as discussed in the preceding paragraphs. Since most prior studies on sensemaking have focused on introductory physics, this study in the context of solving the MoI problems, extends those studies, like another study in the context of Laplace's equation [29], to advance our understanding of the sensemaking of physics graduate students. In particular, Ref. [29] explores in detail students' sensemaking processes in the context of solving Laplace's equation (LE) to find the potential in a spatial region devoid of charges. The primary focus of Ref. [29] is on finding the potential in a semi-infinite channel bounded on three sides, a widely used standard boundary value problem that appears in one form or another in most upper-level electricity and magnetism courses. The sensemaking of graduate students was examined in detail as they tried to find the potential in the region of interest. For example, the thinking that prompts them to attempt (inappropriately) to use the method of images is described in Ref. [29]. In other problems, students were prompted to explain whether solving Laplace's equation would be most appropriate to find the potential in several scenarios [29]. The sensemaking also illustrated how resources such as curving or curling of electric field lines could be used (inappropriately) by some students [29]. Furthermore, several students' sensemaking as they attempted to sketch the equipotential surfaces and electric field lines within the semi-infinite channel was described [29]. The research discussed in this paper has a different focus, as the semi-infinite channel scenario is absent, and only those regions that contain at least one charged particle are considered. In particular, the problems discussed here can be effectively addressed using the method of images, a method that was not appropriate to solve problems in the previous study [29]. Here, we examine students' sensemaking as they solve problems in which the method of images is an effective approach.

Since the method of images is appropriate in the contexts discussed here, students' sensemaking shows activation of a wide variety of related knowledge resources and inferences from them; the variety and depth of ideas they employ to apply the MoI have never previously been explored in literature. Although the framework of "epistemic games" is applied to the analysis of students' sensemaking in both Ref. [29] and the study discussed in this paper, the specific physical contexts, student ideas, and detailed reasoning processes are totally distinct. It is useful for instructors and education researchers to know about how advanced students go about solving problems related to MoI or LE, what knowledge gets activated, what does not get activated, and what happens when you give students a small nudge (e.g., provide a little hint while problem-solving, etc.). Thus, the MoI and LE [29] studies investigate advanced student sensemaking in totally new contexts. The MoI research discussed here leads to novel findings that do not feature in the previous publication on sensemaking in the context of LE [29], relating to two main areas: in-depth comparison between multiple students and in-depth comparison between related but distinct problems within individual students. We discuss student sensemaking across several cases to give deeper insight into graduate students' sensemaking and to compare their sensemaking across different students and cases. The study presented here provides insights into how students transfer and translate strategies between different problems in which MoI is effective and how they use reasoning primitives while sensemaking. The paper presented here significantly elaborates on the Pictorial Analysis game in the 2024 PERC proceedings [71] and provides a detailed analysis of students' sensemaking including other epistemic games.

## II. METHODOLOGY

Similar to our study on advanced students' sensemaking in the context of Laplace's equation [29], this investigation was conducted in the context of collecting qualitative data for the development and validation of a tutorial (along with the corresponding pretest and posttest) on the MoI for the upper-level undergraduate course. We conducted and recorded interviews with six grad students from a large public university in the United States online via Zoom. We interviewed physics grad students who were in different years, ranging from first to fourth year of their research studies, and who had learned these concepts both at the upper-level undergraduate and graduate levels. In particular, they were all familiar with the concept of MoI and had learned it within a few years, including some of them having learned it within the last few months. The MoI was discussed in their upper-level undergraduate coursework and revisited at the graduate level for the students interviewed. We did not find significant differences in students' sensemaking patterns depending on how much time had lagged between their second time learning of this content as graduate students and the interviews. The goal of this investigation is not to follow the progression in sense-making from undergraduate to graduate levels. Instead, grad students' sensemaking is insightful in its own right for understanding how they activate knowledge resources, how they evolve with successive attempts at the same or different problems related to MoI, and how scaffolding impacts their evolution, particularly because the graduate-level physics course in electricity and magnetismalso includes the types of problems discussed here. We did not interview the undergraduate students who were enrolled in the upper-level electricity and magnetism course at that time (when the grad students were interviewed) because they had not had instruction in relevant concepts. However, what we learned from the sensemaking of grad students and what type of knowledge resources they activated while problem-solving can also be valuable for helping upper-level students do expert-like sensemaking. During interviews, students shared their screens on Zoom while working on the problem set given to them in the pdf format. They used an electronic device (iPad) to scribble on the pdf while sharing their screen. They were asked to turn the video off and replace their names with pseudonyms to keep their identities anonymous. At the end of each interview session, students shared their scribbled pdf files with the interviewer. These interviews were conducted for a total average time of approximately 3 h with each student.

The interviews were divided into three sessions, which were conducted on different days of a given week based on the availability of the interviewer and interviewee, or two sessions were merged in some cases if a student was willing to sit longer. Pretest and posttest were given in the unscaffolded version, followed by the scaffolded version for the same problem. An unscaffolded version has question(s) similar to the textbook problems without guidance, whereas a scaffolded version has questions broken down into multiple parts since some scaffolding is provided to proceed. The first session consisted of the unscaffolded pretest followed by a scaffolded pretest on the MoI. The second session consisted of the tutorial, and the third session consisted of the unscaffolded posttest followed by a scaffolded test on the MoI. The interviews used a semi-structured think-aloud protocol in which students were asked to think out loud as they solved or thought about the solutions of the problems given. The interviewer (first author) acted as an observer (jotting down highlights during these interviews), who did not interrupt or help except when clarification was needed or to ask students to keep trying if they were giving up easily on a problem. Immediately after the data collection, the interviewer wrote down narrative summaries of the sessions. The total duration of interviews conducted across the set of pretest, tutorial and posttest on the MoI with six students was 18 h and 16 min. The interviewer saved the video recording, audio transcripts and pdf files of the students' work and summaries of the sessions for further analysis. These interviews are rich in the depiction of the sensemaking of students while they solve the problem.

We used the epistemic game framework to describe the sensemaking of grad students. The process of determining the cases to focus on was done iteratively by the researchers together to devote time to the instances of sensemaking that were most interesting and valuable pedagogically from the perspective of the researchers. We (the first and second authors) discussed the sensemaking episodes together iteratively to agree upon the local coherence and consistency in students' approaches. We also discussed iteratively until we agreed on the analysis of students' sensemaking based upon the different epistemic games. While the two researchers discussed the cases, the interviewer went back to the transcripts and video recordings many times to analyze the details of the excerpts of those chosen instances. The researchers also discussed and converged on our hypotheses and interpretations of how local coherence is observed in students' sensemaking, even though global coherence may not be achieved.

Here, we focus on only those three of the six grad students interviewed who engaged in deep sensemaking and evolving throughout the process of solving problems in order to gain useful insights into grad students' sensemaking. Deep sensemaking does not imply correct problem-solving or superior performance. Rather, it refers to extensive engagement with the reasoning process, willingness to explore different approaches, and articulation of thinking, all of which allow researchers to observe and analyze the sensemaking process. Based on the discussed features of sensemaking observed by researchers and after multiple iterative discussions, we agreed that G1, G2, and G3 were involved in deep sensemaking while solving problems. The remaining interviewed students either did not show enough engagement with the content to reveal useful sensemaking patterns or were extremely proficient such that there was very little interesting insight to be gained or were engaged in sensemaking that was similar to that of the students discussed here. Also, despite their graduate-level training, G1, G2, and G3 struggled with aspects of these problems, and these struggles, combined with their verbalization of reasoning, provide insight into the challenges students face with the MoI. These episodes are discussed in detail to show what knowledge resources are activated in certain situations and how grad students entered and exited a particular game. These episodes also show how grad students evolved from an incorrect approach to a correct approach with the help of nudging or scaffolding.

Out of these three interview sessions for each grad student, we discuss four problems shown below, which are labeled Q1 – Q4. At the beginning of each problem, in the square bracket, we have written the type of test it was part of. At the end of the question, in the square bracket, we have a short hint for the problem solution for the readers; neither the hints nor the answers were provided to students during the interview. Q1 - Q4 discussed in this paper are given as follows.

Q1. [Unscaffolded pretest] Consider the surface of a grounded conductor as the $xy$ plane. The region $z > 0$ is vacuum. Two-point charges are placed on the $z$-axis: charges '$q$' at $z = a$ and '$-2q$' at $z = b$. Write down the expression for the potential in the region of interest. [Correct number of image charges: two, $-q$ $(0,0,-a)$ and $+2q$ $(0,0,-b)$]

Q2. [Tutorial] Consider two semi-infinite, perpendicular, grounded conducting planes ($x = 0$ plane and $y = 0$ plane). A point charge '$q$' is placed at the coordinates $x = a, y = b, z = 0$ in the first quadrant. a) Draw a sketch of this situation. b) To exploit the MoI for the potential in the quadrant of interest, how many image charges are needed? Explain. c) Draw a sketch of where any image charge(s) should be placed to solve for the potential using the MoI. Include the values and signs of the image charges. d) Check the boundary conditions explicitly for the new problem to see if they match the boundary conditions for the original problem. e) Use the MoI to write down the potential in the quadrant where the original point charge is located. [Correct number of image charges: three, $-q$ $(-a, b, 0)$, $+q$ $(-a, -b, 0)$ and $-q(a, -b, 0)$]

Q3. [Unscaffolded posttest] Consider two grounded conducting planes, both perpendicular to the $xy$ plane, intersecting at an angle of 60° as shown in the Fig. 2. One plane is $y = 0$ and the other makes a 60° angle with respect to it. A point charge '$q$' is placed at the point specified by the coordinates $x = \sqrt{3}a$, $y = a$, $z = 0$ (along the line bisecting the region at 30°) in the sextant of interest. Write down the expression for the potential in the region of interest. [Correct number of image charges: five, $-q(0, 2a, 0), +q(-\sqrt{3}a, a, 0), -q(-\sqrt{3}a, -a, 0), +q(0, 2a, 0), -q(\sqrt{3}a, -a, 0)$].

Q4. [Scaffolded posttest] Consider two grounded conducting planes, both perpendicular to the $xy$ plane, intersecting at an angle of 60° as shown in Fig. 2. One plane is $y = 0$ and the other makes a 60° angle with respect to it. A point charge 'q' is placed at the point specified by the coordinates $x = \sqrt{3}a, \ y = a, \ z = 0$ (along the line bisecting the region at 30°) in the sextant of interest. How many image charges are required to solve this problem using the method of images? Explain. a) six image charges, b) five image charges, c) four images charges, d) three images charges. [Correct number of image charges: b) five image charges]. [Note: Q4 was a part of the scaffolded posttest and therefore was followed by another question that asked to write down the expression for the potential in the region of interest. We have not shown that other question here, but we still consider the final expression for the potential to be the goal for the students while discussing episodes of sensemaking.]

Student sensemaking based on these four problems is discussed in three different cases labeled case A, case B, and case C. These cases follow the chronological order and were part of different sessions. The names of students were changed to keep their identities anonymous. In this paper, we call those students G1, G2, and G3. In case A, we discuss a sensemaking episode of grad student G1, who mixed up different methods while solving the problem. In case B, we discuss episodes of sensemaking for Q2 by students G1, G2, and G3. In case C, again we discuss the sensemaking of all three students G1, G2, and G3 (in the Appendix). These students were randomly given names to maintain anonymity, and the number associated with 'G' does not indicate any order. The excerpts from the interviews are quoted in italics in the results and discussion section. Redundant words such as 'like' and "umm" are removed from the transcript, and we added our additional narrative within quotes in square brackets for better understanding and comprehension by the reader. We include our hypotheses and interpretation of evidence of local coherence in student sensemaking in each case. In the summary section, we discuss some similarities among grad students across different cases and across different students.

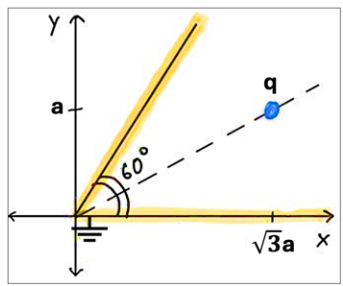

Fig. 2. Picture provided with Q3 and Q4

Here are the research questions that are addressed in this study that we will reflect upon in the Broader Discussion section:

**RQ1:** Which knowledge resources do graduate students activate while solving MoI problems, and how do these resources facilitate or impede problem-solving?

**RQ2:** How do graduate students' approaches to solving MoI problems evolve across different geometric configurations and levels of scaffolding? What are the similarities and differences in the sensemaking of different students?

**RQ3:** What epistemic games do graduate students play when solving method of images problems, and how do these games show nesting and interact during sensemaking?

**RQ4:** How does scaffolding or guidance influence graduate students' sensemaking and progress toward correct solutions?

### III. RESULTS AND DISCUSSION

In this section, we discuss the sensemaking of three grad students G1, G2, and G3, for three cases labeled case A to case C. In case A, we discuss the sensemaking of student G1 for Q1, in which they mixed up different methods while solving the problem and thought that within the method of images (MoI), the purpose of placing image charges was to turn Poisson's equation into Laplace's equation (LE) by making the net charge (including all of the regions) zero. In case B, we discuss the sensemaking of three different students G1, G2, and G3 for Q2 and observe that making diagrams iteratively helps students in improving the problem-solving approach. In case C, we discuss the sensemaking of G1, G2, and G3 (in the Appendix) for Q3 and Q4, which are based on a problem that is reasonably challenging for advanced students and requires them to have a good understanding of the symmetry of the problem. All three cases are discussed in detail in this section through the lens of the epistemic game framework. We discuss evidence of local coherence in student sensemaking, even if their approaches lack global coherence. We also discuss broader implications that we as researchers or instructors should pay attention to in order to provide adequate help to even advanced students so that they can continue to develop expertise in solving physics problems.

A. Mixing up of methods and impact of "nudge"

**Discussion of G1 for case A:** We first discuss the sensemaking of a grad student G1 on question Q1, in which two-point charges are present above a grounded conducting plane, and students were asked to write down the expression for the potential in the region of interest. This problem can be effectively solved using the method of images, in which the grounded conducting plane can be replaced by image charges. Once the transformation of the original problem to the one with image charges happens, writing down an expression for the potential is relatively easy. The potential in the region of interest can be written as the superposition of the potential due to the original charges, and the image charges instead of the grounded conducting plane as given in the original problem. We want advanced students to be able to discern situations in which the method of images is an effective technique, and they should be able to, e.g., use it to find potential. Student sensemaking shows that they not only struggled in recognizing whether the method of images is appropriate in a given situation, but also struggled to determine the correct number, sign, magnitude, and position of the image charges. During the interview, G1 was first provided with problem Q1 in the unscaffolded form, e.g., this problem did not provide an explicit hint about an effective method that could be used to solve the problem. The goal of the problem is to find an expression for the potential in the region of interest.

After reading the problem, G1 began to reflect on it and said, *'…So I like to have a visual in my mind first…'*. The problem prompted them to enter the Pictorial Analysis game. They started drawing axes and the $xy$ plane as shown in Fig. 3, saying, *'…we have the grounded conductor in the $xy$ plane* [drew the conductor] *…this is what we're dealing with… Then we have two-point charges, one of them is 'q' and it's placed at $z = a$* [placed a charge 'q' at $z = a$ as shown in Fig. 3] *… and then we have point charge '-2q' at $z = b$, so we don't have any information about how these 'a' and 'b' are, so I have minus '-2q'* [placed '-2q' charge at distance 'b' on negative z axis]*, and we're trying to find an expression for the potential in the region of interest…'.*

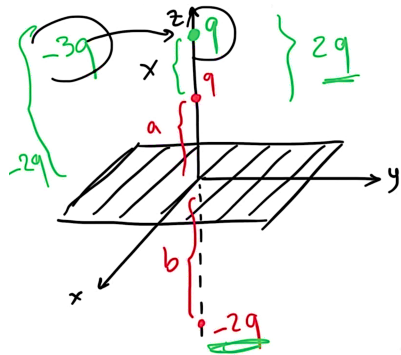

Fig. 3. G1's drawing for Q1

We note that G1 was confused about the position of the original charge '-2q' and incorrectly placed it on the negative z-axis. This incorrect placement of the charge changed the way this problem should be solved as there was no 'excluded region' left where image charges can be placed. Even though G1 was able to recognize that this problem can be effectively solved using the MoI, they were confused about the region of interest. Interestingly, at this point in the problem-solving process, we observed the nesting of the epistemic games when G1 appeared to be entering another epistemic game within the moves of the Pictorial Analysis game. In particular, while figuring out the image charges in the picture, they decided to focus on the mathematical equations and entered the Mapping Mathematics to Meaning game. In this game, their target was to look at the relationship between image charges and a differential equation that they thought was relevant and can help to obtain the correct number and position of the image charges and hence the potential. They said, *'…I'm kind of confused over what we mean by the region of interest. But what I remember is that this is what we had* [wrote down Poisson's equation $\nabla^2 V = -\frac{\rho}{\epsilon_0}$ ] *for solving for potential…over here, we're using method of images…I'm not sure but maybe…we want to add another point charge, so that it kind of balances the system that we're looking at, so that we will turn this equation into 0* [wrote down Laplace's equation, $\nabla^2 V = 0$]. At this point in the problem-solving process, G1 thought that the aim of the MoI is to find image charges such that all charges '*balance*' each other and the net charge becomes zero, and this balancing of charges converts Poisson's equation into Laplace's equation. This idea is not correct because the aim of the MoI is to find image charges such that an easier equivalent problem can be used to obtain the same solution in the region of interest. G1 was interpreting 'balancing' as the process of making the net charge zero by adding the image charges. They misinterpreted the local charge density $\rho$ in Poisson's equation as the total charge, including all regions in the problem, and started looking for image charges such that all charges, including the original charges, sum up to zero. They thought that this would set $\rho = 0$ in Poisson's equation, $\nabla^2 V = -\frac{\rho}{\epsilon_0}$, and it will be transformed into Laplace's equation $\nabla^2 V = 0$, which will need to be solved for the potential.

Once they (incorrectly) decided how the image charges should be placed in the picture (Fig. 3) using the Mapping Mathematics to Meaning game, they resumed the Pictorial Analysis game with the goal of placing the image charges in the picture to 'balance' the charges saying, '*…I think this was what we're trying to do with the method of images. So maybe we could try adding a point charge somewhere over here on the z axis* [placed a point charge 'q' on the positive z-axis as shown in Fig. 3]*…so that we're going to end up like with '2q'* [sum of original charge and image charge in the region above $xy$ plane] *on this side, and then we have '-2q' on the other side* [original charge below $xy$ the plane]. *But I don't know how I'm supposed to find this distance* [distance between the original charge 'q' and the image charge 'q' on the positive z-axis] *…Also,… maybe…instead of 'q'* [the image charge on +z axis]*, I should have added '-3q'* [as the image charge on +z axis]*, so on both sides, I'm going to end up with '-2q'* [as the sum of charges in that region]*'.*

We note that G1 knew that the MoI is an efficient technique that can be used for this problem, and they also knew that there should be some kind of image charges which would help in writing the expression for potential. They used the word 'balance' to figure out the image charges and could not remember that checking the boundary condition is useful to get the correct number, sign and position of the image charges. They focused on creating image charges and toyed with two different configurations of the image charges. In one configuration (see Fig. 3), they placed a +q image charge on the positive z axis to make the total charge (including the original charge) along the positive and negative z axis equal and opposite (i.e., +2q along the positive z axis and -2q along the negative z axis). In the other configuration, they placed a -3q image charge on the positive z axis to make the total charge along the positive and negative z axis equal to -2q as shown in Fig. 3. They finally decided to consider '-3q' image charge on the +z axis and made an implicit assumption that since the -2q charge below the origin was along the negative z axis, the net charge would be $-3q + q - (-2q) = 0$. Once G1 was satisfied with the image charges, they exited the Pictorial Analysis game. They decided to turn in the solution saying, '*once we add that* [image charge they had placed already] *then I think the equation that expresses the potential would be the Laplace equation…*'.

The interviewer stayed as an observer until this point but decided to intervene as G1 was turning in their solution for this problem without solving for the final expression for the potential. The interviewer asked them to think about the expression for the potential by saying, '*…Okay, so do you want to try writing the potential in the region? Want to think about it?'*. This prompt helped G1 take another look at what they had done thus far and reset their approach completely, prompting them to enter the Transliteration to Mathematics game. They started to question whether Laplace's equation (LE) was needed to find the potential in this problem and recalled the problems involving writing the expression for the potential due to multiple point charges they might have solved in the past. They felt that an expression for $V$ can be written without solving LE explicitly here. They then focused their attention back on the picture they had obtained from their Pictorial Analysis game as shown in Fig. 3 and said, '*…I don't know what I'm supposed to be writing. I know that we had to solve for this equation* [Laplace's equation, $\nabla^2 V = 0$] *somehow but I'm not sure how…I think, I remember some stuff. We would write the potential based on the charges that we had* [started writing down the expression for potential]. *So, we have this '-2q'* [original charge on the negative z axis] *over the distance from this surface over here* [distance from $xy$ plane to the charge] *… 'b' plus…depending on either, if I choose this or this* ['q' or '-3q' as the image charge on the positive z-axis]*, which I don't know. Let's go with this one* ['-3q' as the image charge] *-3q over this distance* [distance between $xy$ plane and image charge on +z axis] *let's call it 'x'* [distance between original charge and image charge on +z axis] *plus 'a'* [distance between origin and original charge 'q' on +z axis] *plus 'q' over 'a', I don't know maybe something like this...'*. They wrote the expression for the potential as $V = -\frac{2q}{a} + \left(-\frac{3q}{x+a}\right) + \frac{q}{a}$. We note that G1 did not write the distances in the expression for the potential correctly. The distance should be measured from each charge to the point where the potential is to be calculated. It is possible that they misinterpreted the 'distance' in their formula as the distance of the image charges from the surface. We also note that in this case, G1 did not check the boundary conditions to verify whether their expression for the potential V is correct or not. Fig. 4 shows the nested structure of epistemic games played by G1 while solving Q1. Table 1 shows the summary of sensemaking and Table 2 shows the key sensemaking patterns along with epistemic games for G1 in case A.

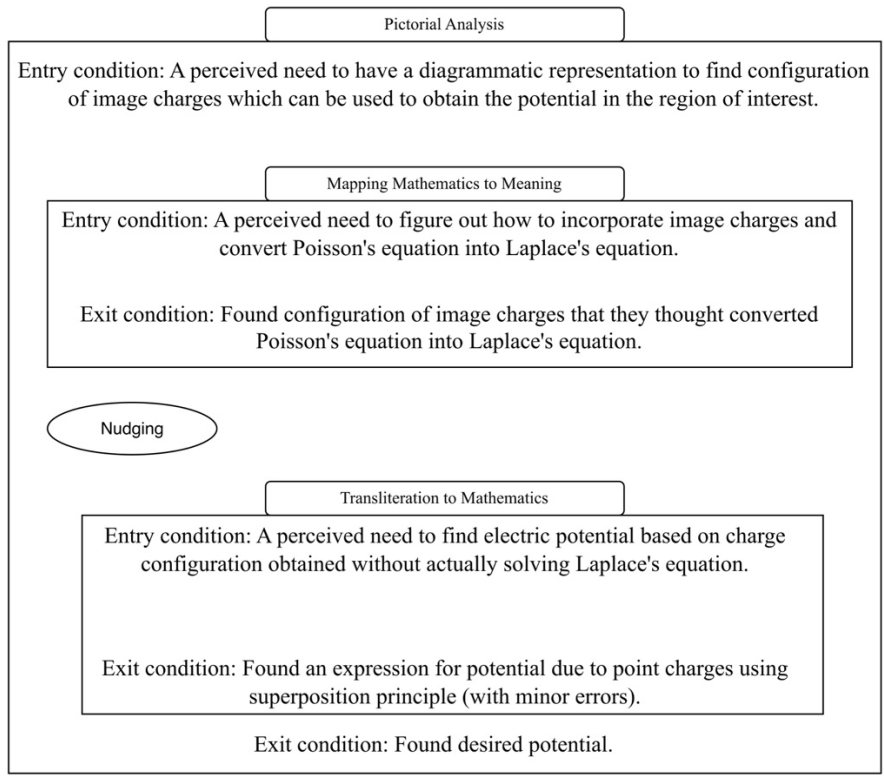

Fig. 4. The nested structure of epistemic games played by G1 while solving Q1

Table 1. Summary of sensemaking of G1 in case A

| Aspect | Summary of G1's Sensemaking in Case A |
|---|---|
| Use of diagrammatic representation | Entered Pictorial Analysis game; drew a diagram showing the grounded conductor and original charges. |
| Charge placement based on misinterpretation of problem statement | Interpreted the problem statement differently and placed -2q charge on the negative z-axis; did not realize that there is no excluded region to place image charges. |
| Nested epistemic games | Entered Mapping Mathematics to Meaning game within the moves of Pictorial Analysis game; used Poisson's equation to reason about the placement of image charges; attempted to "balance" charges such that the net charge, including all regions, becomes zero. |
| Mixing of methods | Thought that turning Poisson's equation (PE) into Laplace's equation (LE) is part of the method of images; thought that after "balancing" charges, LE should be solved to obtain the potential. |
| Confusion between concepts (e.g., local charge density and net charge) | Could not differentiate between local charge density ($\rho$) and total charge (sum of original and image charges across all regions); used idea of "balancing" charges to convert PE into LE. |
| Effect of nudging / interviewer's prompts | Prompting to write potential helped student abandon prior approach of converting PE to LE using image charges; student reset their approach and realized potential can be written without solving LE. |
| Boundary conditions not checked | Did not activate knowledge resources about checking boundary conditions; did not recognize that there should be no image charges in the region of interest. |

| Struggle with distance interpretation in potential | Wrote distances from origin to point charges rather than from a general point in the region (misinterpretation of distances is common among other students discussed here). |
| Persistence of "balancing" idea | Idea of balancing charges carried over and activated later in another context as "equilibrium of forces" (discussed later in case C). |

Table 2. Key sensemaking patterns and primary epistemic games played by G1 in case A

| Student | Key Sensemaking Patterns | Primary Epistemic games (nested) |
| --- | --- | --- |
| G1 | Use of diagrammatic representation.<br><br>Charge placement based on misinterpretation of problem statement.<br><br>Nested epistemic games.<br><br>Mixing of method.<br><br>Confusion between concepts (local charge density and net charge).<br><br>Effects of nudging / interviewer's prompts.<br><br>Boundary conditions not checked.<br><br>Struggle with distance interpretation in potential.<br><br>Persistence of "balancing" idea. | Q1<br>Pictorial Analysis<br>    Mapping Meaning to Mathematics<br>(Interviewer's nudging)<br>    Transliteration to Mathematics |

## B. Effective representation

In this case, we discuss the sensemaking of three students G1, G3, and G2, for Q2, which is a scaffolded problem. In this problem, a charge is present between two perpendicular, grounded, semi-infinite, conducting planes and students are asked to write down an expression for the potential in the region of interest. This problem is slightly more complicated than the previous problem (Q1) discussed in case A, as it includes two conductors and requires students to explore the geometry of the problem. In this problem, for the given charge '$q$', conductors can be replaced by three image charges and the potential can be written as the superposition of the potential due to the original charge and the three image charges. Student sensemaking shows that they struggle less in finding the correct positions of the image charges and more in obtaining the correct number of image charges, as well as the magnitude and sign of those image charges.

**Discussion of G1 for case B:** We begin with the discussion of an episode of student G1 that shows how student sensemaking evolved in the problem-solving process throughout. G1 started by reading the problem carefully. As Q2 a) explicitly asks students to draw a sketch of the situation, we see that G1 immediately entered the Pictorial Analysis game. They started by drawing the axes and the grounded conducting semi-infinite planes and placed the original charge at $x = a, y = b$, as shown in Fig. 5.

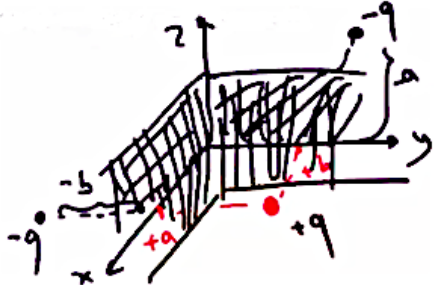

Fig. 5. G1's drawing for Q2

In the next step, Q2 b) explicitly asks students to write down the number of image charges required for the given problem along with an explanation. After reading Q2 b), G1 said, *'…so, this is the region of interest for us* [referring to the first quadrant in Fig. 5]. *I can put a '-q' charge with distance 'b' over here* [placed an image charge '-q' in the fourth quadrant at $x = a, y = -b$ as shown in Fig. 5] *and then I can put another '-q' charge at distance '-a' over here and '-b'* [placed an image charge '$-q$' in the second quadrant at $x = -a, y = b$ in Fig. 5]…*'*. At this point, G1 was not sure if there should be two image charges or three, but then decided to go with two image charges and tried to verify that. They said, *'…let's see if…these two* [image charges] *are enough. I remember sometimes we would add like a third one* [image charge] *on this part* [in the third quadrant] *but I'm not sure…for now let's say two…'*. Within the moves of the Pictorial Analysis game, G1 entered the Physical Mechanism game to make sure that considering two image charges would be appropriate. G1 thought that this problem can be thought of as a superposition of two different problems with each problem containing a charge present near a grounded conducting plane at $x = 0$ or at $y = 0$. They said, *'…It's like a mixture of what we had before* [referring to other problems they had solved before this problem with a charge present near an infinite grounded conducting plane]…*the conductor plane question. So I can look at them separately instead and say that I only have the $x = 0$ plane as my grounded conducting plane. So I would need a '-q'* [charge] *for that one with the same distance. And then I also have the $y = 0$ plane separately. So again, I would need to add a '-q'* [charge] *to…solve for that.'* As G1 had already shown image charges in Q2 a) as shown in Fig. 5, they pointed to that same figure without drawing it for Q2 c). They moved on to the next question Q2 d) and started checking the boundary conditions by comparing the original problem given in Q2 with a charge and two perpendicular conductors to another problem where the conductors are replaced by the image charges. They started writing down the boundary conditions for the original problem playing the Mapping Meaning to Mathematics epistemic game saying *'…so for the original problem at $x = 0$, we had the conducting plane, so potential is 0* [ wrote down: $x = 0 \rightarrow V = 0$], *at $y = 0$ again, we have the potential to be 0* [wrote down: $y = 0 \rightarrow V = 0$], *and then at infinity for x goes to infinity, again we have potential to be 0* [wrote down: $x \rightarrow \infty \rightarrow V = 0$]… *y goes to infinity [is] the same* [wrote down: $y \rightarrow \infty \rightarrow V = 0$]… *So as z goes to infinity in our region of interest…we do have semi-infinite planes…that would mean that it doesn't matter…how far I choose my point to be…because the planes are infinite. That would still mean that I'm going to get the zero potential. So…for even z goes to infinity, I still have that equal to 0* [wrote down: $z \rightarrow \infty \rightarrow V = 0$]*'*. G1 then started focusing on the boundary conditions for another ('new') problem (without conductors but with image charges). They said, *'Now for the new problem…I don't have the conducting planes anymore, but I have the point charges to look at…'*.

After this, they resumed the Pictorial Analysis game and started drawing a new picture as shown in Fig. 6, saying, *'…so I have something like this* [drew the axes as shown in Fig. 6]. *I have the '+q' over here* [placed original charge '+q' in the first quadrant as shown in Fig. 6], *I have a '-q' over here* [placed image charge '$-q$' in the fourth quadrant], *and another '$-q$' over here* [placed another image charge '$-q$' in the second quadrant]*.'*

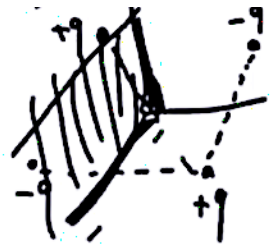

Fig. 6. G1's drawing for Q2 d) while looking at the boundary condition

At this point, G1 was confused as this problem has semi-infinite plane whereas they were considering the superposition of two infinite planes. They said, *'...So by doing this, the potential at this surface again will be 0* [referring to $x = 0$ axis]. *I'm doubting on that, but also for this one, because I don't have an infinite-infinite plane from this side* [it is semi-infinite instead of infinite, $x = 0$ plane]. *It says it's semi-infinite, so I don't know if that would play a role, if it went all the way to here* [extended towards $-x$ axis]. *Maybe I could say that but it's not. So maybe I need to do something else.'* G1 then decided to add another image charge in the third quadrant, *'Let's say I added… this image* [charge] *'q' maybe over here* [added third image charge 'q' in the third quadrant in Fig. 6] *...'*.

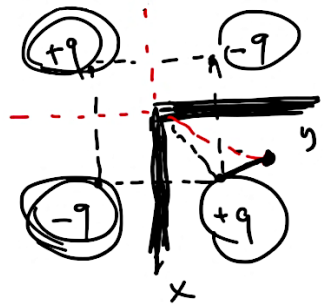

Fig. 7. G1's drawing for finding the number of image charges in Q2

Within the Pictorial Analysis game, G1 then decided to draw another picture without showing planes in 3D space, as shown in Fig. 7 saying, *'...if I don't have this plane…I'm just looking at a system of charges $+q, +q, -q, -q$* [drew four charges at four corners of an imaginary square, as shown in Fig. 7] *… So originally, I am looking at this problem. So I have my conducting planes over here* [drew the semi-infinite conducting planes using bold lines as shown in Fig. 7]. *So, I want my potential to be 0 on this surface* [$x = 0$ plane] *and on this surface* [$y = 0$ plane] *…also, if I don't add this at this '+q' over here* [image charge in the third quadrant of $xy$ plane in Fig. 7]. *I'm not sure if it would still satisfy the boundary conditions because the planes are not infinite... If the planes were infinite, then I guess it would make sense. But since they stop at this point* [along $z$-axis]…*I guess it would make sense…if I have…imaginary planes going this way* [extended planes on the other side of the axes to show imaginary planes as shown in Fig. 7], *then for each of them, I could say, okay, now I have '+q', '-q',…'-q', '+q'. I think the symmetry makes more sense for me like this. So let's say I change this to three* [answer of Q2 b) from two to three] *…At this point over here in the back I'm also gonna add a '+q' over here* [added the third image charge in the third quadrant in Fig. 5, which now looks like Fig. 8. *Yeah, I think this way if I extend these planes, not really but if we imagined that these planes were extended* [in Fig. 7, student points to the imaginary extended planes], *then all of the boundary conditions would be satisfied.'*

With this thought in mind, at this point, they jotted down the expression for the potential as $V = \frac{1}{4\pi\epsilon_0}\left(\frac{q}{\sqrt{x^2+y^2+z^2}} + \frac{-q}{\sqrt{(x+a)^2+y^2+z^2}} + \frac{-q}{\sqrt{x^2+(y+b)^2+z^2}} + \frac{q}{(x+a)^2+(y+b)^2+z^2}\right)$. In this case, G1 was eventually able to figure out the correct number of image charges at the correct position but we note that they made minor errors in writing the distances in the expression for the potential.

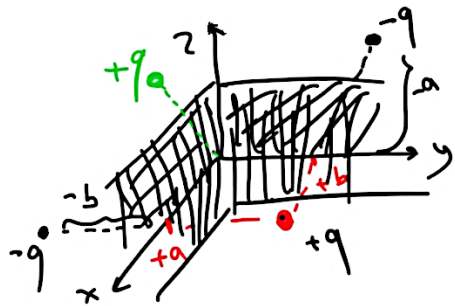

Fig. 8. G1's drawing after adding third image charge in their first picture (Fig. 5) for Q2

Fig. 9 shows the nested structure of epistemic games played by G1 while solving Q2. Table 3 shows the summary of sensemaking of G1 in case B.

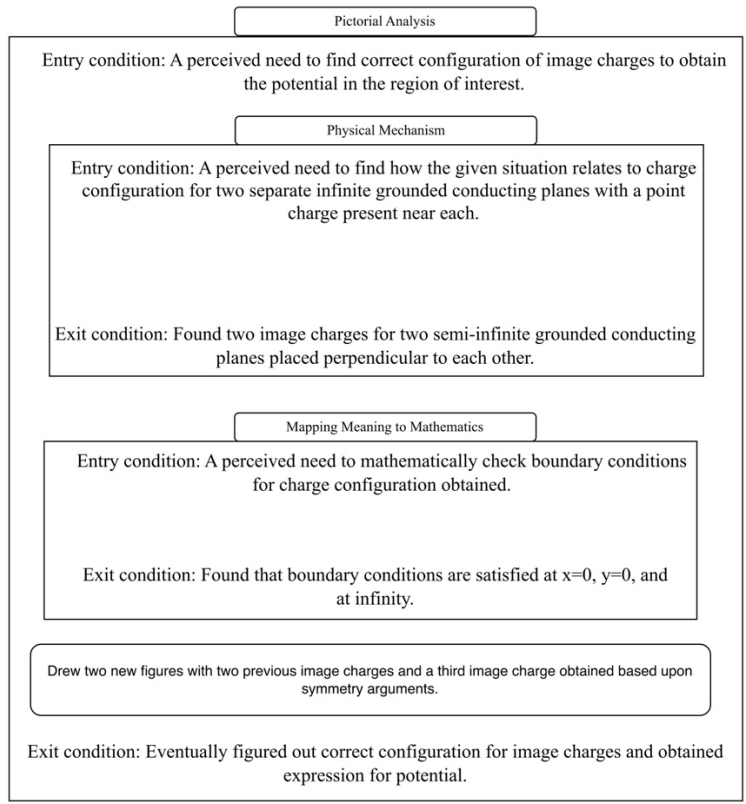

Fig. 9. The nested structure of epistemic games played by G1 while solving Q2

Table 3. Summary of sensemaking of G1 in case B

| Aspect | Summary of G1's Sensemaking in Case B |
|---|---|
| Use of diagrammatic representation | Entered Pictorial Analysis game; drew coordinate axes and planes and placed original charge at ($x = a$, $y = b$). |
| Initial assumption about number of image charges based on two planes | Initially placed two image charges ($-q$ at $x = a$, $y = -b$ and $-q$ at $x = -a$, $y = b$); uncertain about need for a third image charge; decided to go with two image charges. |
| Nested epistemic games | Within Pictorial Analysis game, entered Physical Mechanism game and then Mapping Meaning to Mathematics game. |
| Use of superposition of two different planes | Applied superposition reasoning; treated the problem as two independent cases of a charge near a single infinite grounded conducting plane, using prior experience with similar problems. |
| Checked boundary conditions | Identified and checked boundary conditions for the original problem: $V = 0$ at $x = 0$ and $y = 0$, and $V \to 0$ as $x, y, z \to \infty$; thought boundary conditions were satisfied with only two image charges. |
| Confusion between semi-infinite and infinite planes | Confused between semi-infinite and infinite grounded conducting planes; questioned whether semi-infinite nature affects potential and boundary conditions. |
| Refinement of model | Added a third image charge ($+q$ in the third quadrant) to account for the semi-infinite planes; redrew figure with four charges ($+q$, $-q$, $-q$, $+q$) placed symmetrically with respect to imaginary extended planes. |

| Iterative refinement of figures | Drew multiple figures at different viewing angles to understand the symmetry of the problem. Progressed from guessing and verification to refinement through iterative diagrammatic reasoning and boundary condition analysis. Eventually arrived at an almost correct configuration and potential expression. |
|---|---|

**Discussion of G3 for case B:** We now discuss the case of another student, G3, who attempted to solve problem Q2. G3 started by reading the problem carefully. Then, for Q2 a), they entered the Pictorial Analysis game and started drawing the axes as shown in Fig. 10 saying, *'I'm trying to draw this in a good way in 3D. I'll draw in 3D and if it doesn't work, then I'll draw it in 2D...That is my y [y = 0] plane [drew y = 0 plane as shown in Fig. 10] ...That's my x =0 plane [ drew $x = 0$ plane]*. G3 then placed the original point charge in the first quadrant saying, *'I don't know what 'a' and 'b' are relative to each other but I'll just say that 'b' is greater than 'a' for now...my point charge...[is at]...a, b and 0 [placed the original charge at (a, b, 0)]'*. G3 then moved to Q2 b) to figure out the correct number of image charges. Within the moves of the Pictorial Analysis, they started playing the Physical Mechanism game to figure out the relationship between the number of image charges and the number of excluded regions (G3 incorrectly thought that there were multiple excluded regions).They said, *'...I don't know if it's dependent on the actual number of conductors, the conducting surfaces, or the charges...we can look at the two different regions to calculate the potential, the positive x region [wrote down: $x > 0$] and the positive y region [wrote down: $y > 0$]...I don't know if that means that we need one image charge to be able to calculate ... the potential at both those regions in the excluded region or if we need two separate charges to calculate these potentials because they are two different planes.'*

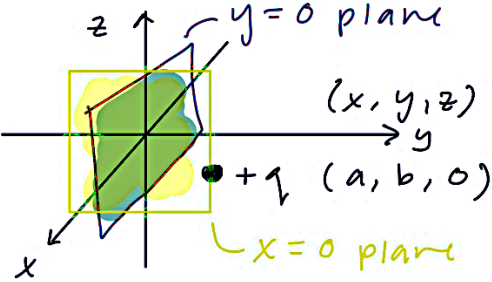

Fig. 10. Picture drawn by G3 to answer Q2 a)

G3 felt that they should redraw the picture and hence resumed the Pictorial Analysis game saying, *'...I didn't really draw this nicely...I'm gonna try drawing this in 2D, maybe it will...help me think about this better.'* G3 started drawing a new picture, as shown in Fig. 11, drew the planes and placed the original charge in the first quadrant, and labeled all the quadrants as I to IV. Although they mentioned '2D' but they continued drawing '3D' (looking) figure in Fig. 11. Then they started thinking about the number of image charges that should be required, saying, *'...Okay, so we're in one of the quadrants [as shown in Fig. 11]. That's fine. Why would I need...more than one though? I don't know. I don't really have a reason for more than one. Oh, wait, okay, if the quadrant of interest is I, then II, III, and IV are considered the excluded regions, right?...Excluded regions. We need three image charges. I think I'll say 3 for now because I'll consider those quadrants [II, III, and IV] as the excluded regions'*.

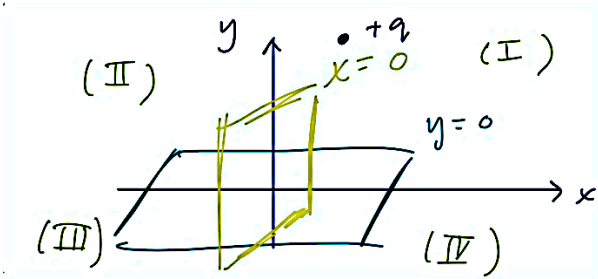

Fig. 11. Picture drawn by G3 to figure out the excluded regions and number of image charges for Q2

After this, they started reading Q2 c) and decided to draw another figure as shown in Fig. 12 without the conducting planes, saying, *'...I'm just gonna draw this as a quadrant for now because now I think I can technically disregard the plane conductors…'.* They placed the original charge in the first quadrant at $(a, b, 0)$ as shown in Fig. 12 and said, *'…I don't know*

*which ones are gonna be positive or negative, though… I'll make them all negative for now* [they placed all other image charges: '-q' at *(-a,b,0)*, '-q' at *(-a,-b,0)* and '-q' at *(a,-b,0)* [later these charges were made positive by G3]]*…I don't really have a good explanation…'.* They thought that they could look at the induced charges on the surface to figure out the sign of these image charges, but they did not consider the fact that the planes are grounded.

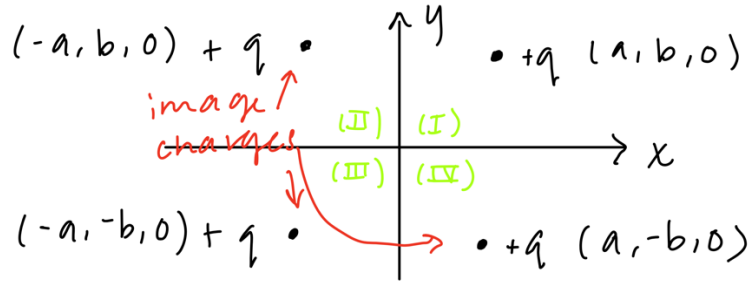

Fig. 12. Picture drawn by G3 to show the image charges for the given problem for Q2

They started drawing the planes to show the induced charges separated on the planes as shown in Fig. 13 and said, *'…I need to look at the induced charges on the surfaces of the planes…so looking at the $x = 0$ plane* [drew a vertical plane as shown in Fig. 13]*, there's going to be…positive to the left* [side of the conducting plane]*…negative to the right* [side of the conducting plane]…[they ignored the grounding of these conductors]*…So this was the induced charge 'q' induced…Okay,* [I am] *gonna look at the other planes separately* [drew another plane referring to y = 0 plane as shown in Fig. 13]*, negative charges right here* [top of the plane]*, positive charges down here* [on the lower side of the horizontal plane as shown in Fig. 13]*. Are they all going to be positive?...Hmm, okay, I'll just stick with negative…image charges just because in the other example that we went through, they were the image negative charges….Oh, wait, now, okay….I'll just have them all as positive* [sign of image charges] *for now. Just because if I do, the induced charges separate on the planes, then there are negative* [pointing to induced charges on one side of the plane] *and there're positive* [induced charges]*. I don't think they should all be the same though.'* Once they figured out the correct number of image charges, they entered the Mapping meaning to Mathematics game. They read Q2d), and they started writing down the boundary conditions, saying, *'…So when I was far away, potential was 0* [wrote down*: $V(r \to \infty) \to 0$]…the other one is at the surface* [wrote down: $V(x, y = 0) = 0$]*, on the surface of the conductor, it's 0.'* After that, they read Q2e) and wrote down the expression for the potential as follows:

They wrote $r_1 = \sqrt{a^2 + b^2}$, $r_1 = r_2 = r_3 = r_4$. $V_{TOT} = V_I + V_{II} + V_{III} + V_{IV} = \frac{kq}{r_1} + \frac{kq}{r_2} + \frac{kq}{r_3} + \frac{kq}{r_4}$

$V_{TOT} = \frac{4kq}{r_1} = \frac{q}{\pi\epsilon_0}\left[\frac{1}{\sqrt{a^2+b^2}}\right]$. They did not write the expression in terms of coordinates $x, y$, and $z$ and hence, the expression written by them is not a general expression for the potential. Summary of sensemaking of G3 in case B is discussed in Table 4.

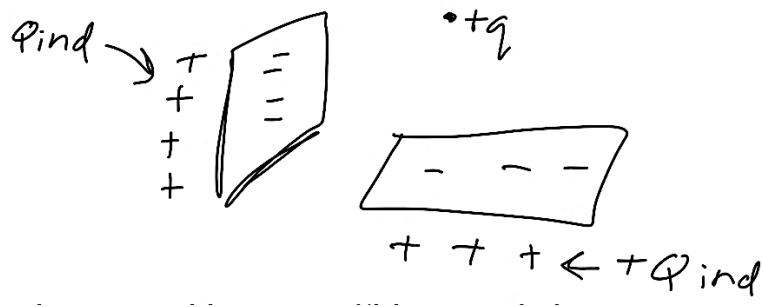

Fig. 13. Picture drawn by G3 to show the charge separation on the planes for Q2

Table 4. Summary of sensemaking of G3 in case B

| Aspect | Summary of G3's Sensemaking in Case B |
|---|---|
| Use of diagrammatic representation | Entered Pictorial Analysis game; tried to draw diagram at the "best" viewing angle. |

| Nested epistemic games | Entered Physical Mechanism game within moves of Pictorial Analysis game to find relation between number of image charges and excluded regions. |
|---|---|
| Search for patterns or logic for number of image charges | Tried to find if there is relation between number of image charges and number of conductors, conducting surfaces, given charges, or number of excluded regions. |
| Iterative refinement of figures | Drew multiple figures from different perspectives to understand the symmetry of problem in 3D. Each drawing revision contributed to better spatial reasoning. |
| Argued for charge separation on the conductors despite grounding | Interpreted "grounded" as "neutral" rather than "zero potential"; represented charge separation on conductor surface with negative charge induced near the original charge and positive charge on the far side; maintained induced charge separation even with grounding. |
| Correct image charge configuration despite inconsistent reasoning | Obtained correct image charge configuration based on symmetry even though the reasoning was not completely correct. |
| Struggle with distance interpretation in potential | Treated all distances as equal and measured from origin to each charge rather than from a general point in the region of interest where potential is to be found. |
| Persistent idea of charge separation on grounded conducting plane and search for logic | Carried along the idea of charge separation on the conductor despite grounding and searched for rationale to find relationship between image charges and "number of excluded regions" (discussed later in case C). |

**Discussion of G2 for case B:** We now discuss the sensemaking of G2 for Q2. G2 started solving Q2 a) by drawing $x, y, z$ axes as shown in Fig. 14 (a) and entered the Pictorial Analysis game. They said, *'...so $x = 0$ would be* [drew $x = 0$ plane and labeled $V = 0$ as shown in Fig. 14 (a)] *...this here...and then $y = 0$ would be this plane* [drew y = 0 plane and labeled $V = 0$]*...and a point charge is placed at the coordinates $x = a, y = b, z = 0$* [placed the point charge '+q' in the first quadrant]*...'*. After they moved to Q2 b), which asks students to write down the number of image charges required along with their reasoning, they said, *'...we have a charge and it is somewhere not on an axis and I want to say we would need two* [image charges] *and now I'm gonna try to justify that...'*. G2 then started thinking about the image charges that should be placed and said, *'...I'm thinking of putting the images over here* [placed an image charge in the fourth quadrant at $x = a, y = -b, z = 0$] *and over here* [placed another image charge in the second quadrant at $x = -a, y = b, z = 0$]*...I'll try to make it make sense...'*. G2 then decided to skip this part of the problem and started working on Q2e) to write down the potential first and then check the boundary conditions to justify the choice of two charges and entered the Mapping Meaning to Mathematics game. They wrote down the expression for the potential as $V = \frac{kq}{\sqrt{(x-a)^2+(y-b)^2+z^2}} - \frac{kq}{\sqrt{(x-a)^2+(y+b)^2+z^2}} - \frac{kq}{\sqrt{(x+a)^2+(y-b)^2+z^2}}$ and said, *'...and now if I plug in $x = 0$ and $y = 0$ to all of these, I would get* [wrote down: $V = \frac{kq}{\sqrt{a^2+b^2+z^2}} - \frac{kq}{\sqrt{a^2+b^2+z^2}} - \frac{kq}{\sqrt{a^2+b^2+z^2}}$ ]*,...which is not 0 but if I were to make the 'q's half* [in magnitude] *of their value, so if I were to make this half and* [this] *half, that* [potential] *would equal 0...at $V(0,0,z)$. So, at the boundary...*[potential] *would equal 0'*. They then multiplied by $\frac{1}{2}$ the second and third terms in the expression for the potential so that it became $V = \frac{kq}{\sqrt{(x-a)^2+(y-b)^2+z^2}} - \frac{kq}{2\sqrt{(x-a)^2+(y+b)^2+z^2}} - \frac{kq}{2\sqrt{(x+a)^2+(y-b)^2+z^2}}$. After this, in response to Q2 b) they reflected and wrote, *'two image charges, since there are now two finite boundary conditions (the two planes $x = 0$ and $y = 0$)'*. While answering Q2 c), G2 decided to draw the picture again, saying, *'...I'll make it a little more clear I think...'*. They started by drawing $x, y, z$ axes as shown in Fig. 14 (b) and then placed the original charge '+q' in the first quadrant and the two image charges '$-\frac{q}{2}$' and '$-\frac{q}{2}$' in the second and fourth quadrant. They then checked one more boundary condition, saying, *'...as x and y go to infinity, V also goes to 0* [wrote down $x, y \rightarrow \infty, V = 0$]. *So that also supports the original boundary conditions'*. Summary of sensemaking of G2 in case B is discussed in Table 5.

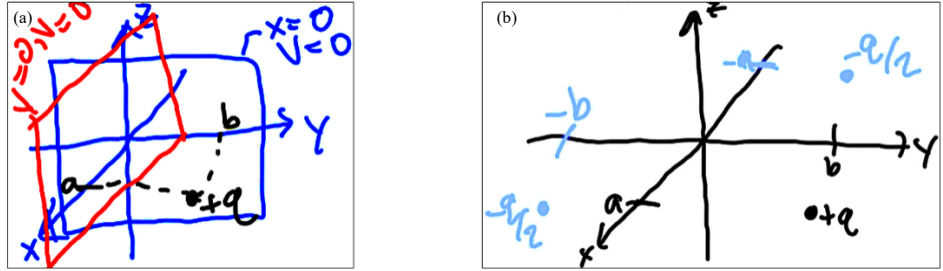

Fig. 14. (a) G2's drawing for Q2 a) and Q2 b) (left), (b) G2's drawing for Q2 c) (right)

Table 5. Summary of sensemaking of G2 in case B

| Aspect | Summary of G2's Sensemaking in Case B |
|---|---|
| Use of diagrammatic representation | Entered Pictorial Analysis game; drew the relevant coordinate axes and conducting planes and placed original charge +q in the first quadrant at (a, b, 0). |
| Initial intuition that there are two (instead of three) image charges due to reflection from two conducting planes | Assumed two image charges (there were two conducting planes) should be required and worked to justify the assumption. |
| Nested epistemic games | Entered Mapping Meaning to Mathematics game within Pictorial Analysis game to check the boundary conditions (b.c.) at $x = 0$ and $y = 0$; observed nonzero potential at the boundaries so b.c. not satisfied. |
| Two image charges with magnitude q/2 (instead of 3 with magnitude q) | Adjusted magnitudes of image charges to half to force $V = 0$ on these boundaries. |
| Iterative refinement of figures | Redrew figure to clarify placement of original and image charges; updated diagram with '–q/2' image charges in second and fourth quadrants. |
| Partial checking of boundary conditions | Checked b.c. at $x = 0$, $y = 0$, and at $x, y \to \infty$; did not check each plane separately; assumed checking b.c. at origin and infinity was sufficient which seemed to be satisfied; assumed potential was obtained correctly. |
| Persistence of the idea of image charge with magnitude q/2 | Carried along the idea of choosing the magnitude of image charges as q/2 to the next problem (discussed later in case C). |

**Case B summary:** Thus, in case B, we discussed the sensemaking of three different students G1, G3, and G2 and noted some common patterns across students. Problem Q2 was scaffolded, and the scaffolding significantly helped students in navigating the sensemaking to solve the problem. Q2 was broken down into multiple parts, and students were asked to draw the picture explicitly, which scaffolded them while sensemaking. They were asked to write down the number of image charges and then show the image charges in a sketch, which often made them focus on more in-depth features of the problem. For example, in problem Q2, there were two sub-problems, Q2 a) and Q2 c), asking students to sketch the situation at two different stages of the problem. We find that students drew more than two pictures, and they got better at identifying the deep features of the problem with every additional attempt. It is possible that if the same problem had been provided in an unscaffolded form, directly asking students to write down the expression for the potential, they might not have made efforts to break the problem into sub-problems with multiple diagrams and would have proceeded with a single and less effective sketch. Thus, scaffolding support can be helpful in guiding student sensemaking while solving these types of problems. With the level of scaffolding problem Q2 provided, students may not end up getting the correct answer, but it is likely to make them more successful in problem solving and learning. We also note that G1 and G3 broke down this problem into two separate problems and then used the superposition principle. G2 implicitly used this concept of considering each conducting plane separately and checked the boundary conditions only at the origin and infinity (did not check it for each plane separately). Writing the distances incorrectly is also a common issue observed in other cases, as well as in the episodes of G1 and G3 for case B. We observed that students carry along the incorrect concepts to other problems. G2 used q/2 as the magnitude of image charge and G3 interpreted the 'grounding' of the conductor as a charge neutral conductor in case C discussed later. The instances discussed earlier suggest that it is important for instructors to provide scaffolding support, e.g., in the form of check points in their instructions to make sure that students do not continue to activate the same knowledge resources that are not appropriate in those situations across multiple problems. Providing adequate scaffolding support to solve the problem at the right time and providing them with time and incentive to reflect upon why some knowledge resources may not be relevant in those contexts could help students learn. These examples also suggest that many problems in physics rely on taking advantage of the pictorial (diagrammatic) and other representations. Therefore, it is very important for the instructors to help students become proficient in multiple representations (diagrammatic, graphical, tabular, verbal, etc.) so that they can activate relevant resources for effective representation while sensemaking in different problem-solving tasks. Key sensemaking patterns and primary epistemic games played by G1, G3, and G2 in case B are discussed in Table 6.

Table 6. Key sensemaking patterns and primary epistemic games played by G1, G3, and G2 in case B

| Student | Key Sensemaking Patterns | Primary Epistemic games (nested) |
|---|---|---|
| G1 | Initial assumption about number of image charges based on two planes.<br><br>Use of superposition of two different planes.<br><br>Checked boundary conditions.<br><br>Confusion between semi-infinite and infinite planes.<br><br>Refinement of model. | Q2<br>Pictorial Analysis<br>    Physical Mechanism<br>    Mapping Meaning to Mathematics |
| G3 | Search for patterns or logic for number of image charges.<br><br>Argued for charge separation on the conductors despite grounding.<br><br>Correct image charge configuration despite inconsistent reasoning.<br><br>Persistence of the idea of charge separation on the plane and search for logic. | Q2<br> Pictorial Analysis<br>    Physical Mechanism<br>    Mapping Meaning to Mathematics |
| G2 | Initial intuition that there are two (instead of three) image charges due to reflection from two conducting planes.<br><br>Two image charges with magnitude q/2 (instead of 3 with magnitude q).<br><br>Partial checking of boundary conditions.<br><br>Persistence of the idea of image charge with magnitude q/2 | Q2<br>Pictorial Analysis<br>    Mapping Meaning to Mathematics |

## C. Exploring symmetry and the impact of scaffolding

Now we discuss the sensemaking of three students - G2, G3, and then G1- for Q3, which is an unscaffolded problem, and Q4, which is a part of the scaffolded version of the same problem. In this problem, a point charge is present in between two semi-infinite, grounded conducting planes making an angle of 60°, and students are asked to find the potential in the region of interest. This is a challenging problem as it requires students to take advantage of the symmetry of the problem. This problem is similar to the problem given in Q2, except that the grounded conducting planes now make a 60° angle instead of 90°. This changes the number of image charges required for the problem from three (in Q2) to five (in Q3). Once image charges are obtained, the correct expression for the potential can be written using the superposition of the potential due to the original point charge and five image charges. For the students who have successfully solved Q2, it is challenging to recognize that this problem requires five image charges rather than three. We find that students struggle in recognizing the symmetry of the problem and finding the correct number of image charges. Obtaining the correct position of all five image charges is also a challenging task for the students.

**Discussion of G2 for case C:** We now discuss the sensemaking of G2 for Q3, for which Fig. 2 was provided to the students along with the question during the interview. G2 read the question carefully and made an educated guess that they would need two image charges, and tried to check the boundary conditions to justify their choice of the image charges. However, G2 struggled to write the boundary condition at the surface of the grounded semi-infinite conductor, making an angle of 60° with the $xz$ plane. Therefore, they left the space empty and wrote down the next boundary condition saying, '...*I'm guessing that we might need two image charges but let's see. So I'm gonna write down the boundary conditions since we already have a picture here* [referring to Fig. 2 provided with Q3].' With this, they wrote the boundary conditions as $V = 0$ at $y = 0$, $V = 0$ at '*this line* [referring to the grounded conducting plane placed at 60°, as shown in Fig. 2], *here that makes...60° angle and so...I can write an equation for that line* [but then they did not write anything]...' and $V = 0$ for far from '$q$'. To be able to visualize the problem more effectively, G2 decided to add details in Fig. 2 that were provided with the problem. They entered the Pictorial Analysis game in which they first labeled what they thought was relevant for the boundary conditions and started adding image charges as shown in Fig. 15.

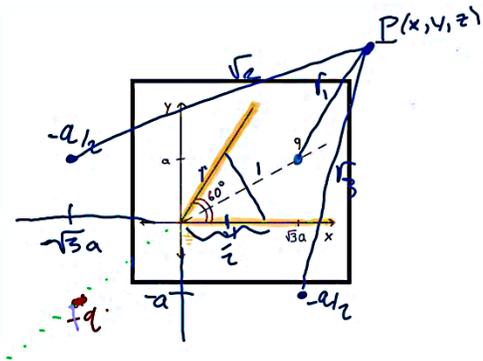

Fig. 15. Picture drawn by G2 while solving Q3

G2 was able to recognize that the problem can be solved using the MoI, and conductors should be replaced by the image charges. However, their sensemaking showed struggle in figuring out the correct positions of those charges. They said, '*if we replace these conducting planes with image charges at specific locations, we would have an equivalent problem. So we need to find where to put these image charges. I guess we could try...mirror locations...at* $-\sqrt{3}a$ [placed a charge in the second quadrant at $x=-\sqrt{3}a$ and $y = a$ as shown in Fig. 15 $y = a$] *and this* [another charge] *would be at minus a* [placed another charge in the fourth quadrant at $x = \sqrt{3}a$ and $y = -a$] *and we're just gonna try that*'. We observe the nesting of epistemic games here such that G2 entered another game, the Transliteration to Mathematics game, within the move of the Pictorial Analysis game. In particular, G2 stated that this problem is similar to problem Q2, which also has a charge present between two conducting semi-infinite grounded planes, and therefore, the charge configurations should be similar in these cases. Earlier we discussed that in case B, G2 incorrectly obtained two image charges with q/2 as the magnitude. It turns out that G2 activated the same knowledge resource and carried along the incorrect concept of magnitude 'q/2' to this new problem, Q3. With this thought in mind, G2 wrote down the magnitude of the charges saying, '*...and from last time* [rereferring to Q2, the problem discussed in case B]...*we needed two image charges, they needed to be half the magnitude. So, I'm gonna...*[put] *opposite sign. So I'm gonna make this* [image charge in second quadrant as shown in Fig. 15] *–q/2 and this one* [image charge in fourth quadrant] *also –q/2 and we'll see how that works...*'. With this, G2 resumed the move of the Pictorial Analysis game and chose a point 'P' in the region of interest where the potential is to be calculated, as shown in Fig. 15.

Now their task was to write an expression for the potential in the region of interest. They also noted that they wanted to make sure that the boundary conditions were satisfied with the image charges they obtained. Thus, at this point, they entered the Mapping Meaning to Mathematics game and started writing an expression for the potential using the original charge and two image charges. They considered the equation for the grounded conducting plane making an angle of 60° with the $y = 0$ plane as $y = 2x$. They spent a significant amount of time substituting boundary values in their expression, but could not get the boundary conditions satisfied with these two image charges (the correct number of image charges required is five). When G2 could not get the desired result, they decided to redraw the picture from scratch, as shown in Fig. 16 and hence resumed the Pictorial Analysis game again. They started by drawing the axes, grounded conducting planes, and placed the original charge. They placed two image charges '-q' at $x = 0, y = a$ and '-q' at $x = 0, y = -a$. They wrote the expression for the potential and checked if the boundary conditions are being satisfied. As G2 did not see this configuration giving the desired result, they extended the grounded conducting planes into the third quadrant using dashed lines and placed the third image charge '-q' at $x = -\sqrt{3}a, y = 0$ to check whether the boundary conditions would be satisfied. As they did not succeed in satisfying the boundary conditions, they felt they could not proceed further and decided to end the game here and turn in their solution to this problem.

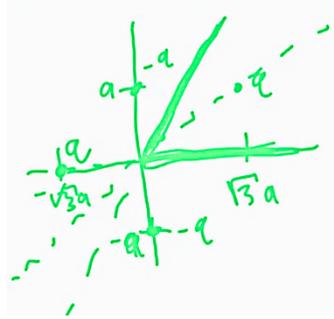

Fig. 16. Picture drawn by G2 during second attempt for Q3

Although G2 could not obtain the correct number of image charges, as well as their positions and values, and hence the correct expression for the potential, we observe an improvement in their approach with multiple iterations in the drawing. We note that students may prefer to draw multiple diagrams for the problem if they are given sufficient time, which can also help them in developing a better understanding of the problem. The problem given in Q3 is an unscaffolded problem, and therefore, students get minimal assistance from the problem statement.

Then, we gave the scaffolded version of the same problem to understand the level of growth in student sensemaking and whether it helps the students in steering them toward the correct answer. One of the problems in the scaffolded version is Q4, in which students were asked to explicitly choose the number of image charges required to solve the problem. Students were provided with the options to choose from possible number of image charges (see Q4). This problem with the prompt gave students an incentive to particularly focus on the number of image charges. This also made them think about why they should be discarding the other options before choosing the right one. We find that it made students think about the symmetry of the problem as they had to think about reasons for both choosing a certain number of image charges and discarding others.

When G2 was given Q4 they said, *'…I was trying three before, six seems excessive. I don't know that we need six…probably need some symmetry…'*. Drawing a diagram is a crucial part of the MoI, and this hint invoked G2 enter the Pictorial Analysis game. They started with Fig. 2, which was already provided in the scaffolded version, and added elements to it to obtain Fig. 17. They extended grounded conducting planes given in Fig. 2 and drew lines to divide the $xy$ plane into six equal regions, as shown in Fig. 17, and said, *' looking for maybe five* [image charges]*, it will look symmetric…maybe I do need one like here* [placed an image charge 'q' diagonally to the original charge in the fourth sextant]*, maybe if I did place them like here* [placed an image charge '-q' in the third sextant]*, here* [placed third image charge '-q' in the sixth sextant]*, so three or five. Maybe four* [image charges] *also doesn't make sense. Let's start with three* [image charges] *and we'll see where that goes…'*. They entered Mapping Meaning to Mathematics and wrote down the boundary conditions as $V = 0$ at $y = 0$, $V = 0$ at $y = \sqrt{3}x$ (after recognizing the geometry) and $V = 0$ far from 'q' (charge). They wrote the expression for the potential using the original charge and three image charges and started checking the boundary conditions. G2 decided to add two more image charges '-q' at $x = 0, y = \sqrt{3}a$ and '-q' at $x = 0, y = -\sqrt{3}a$ as shown in Fig. 17, and started checking the boundary conditions. When the charge configuration did not help in satisfying the boundary conditions, they changed the positions of the recently added image charges from '-q' at $x = 0, y = \sqrt{3}a$ and '-q' at $x = 0, y = -\sqrt{3}a$ to '-q' at $x = 0, y = a$ and '-q' at $x = 0, y = -a$. They were able to figure out that the correct number of image charges should be 'five' as it looked more symmetric with respect to the configuration, but they struggled to find the correct positions of these image charges, and therefore, could not get the boundary conditions satisfied even after spending a significant amount of time. Eventually, they felt they were stuck and decided to end playing the games and reported that 'five' image charges seem reasonable to them. Summary of sensemaking of G2 in case C is discussed in Table 7.

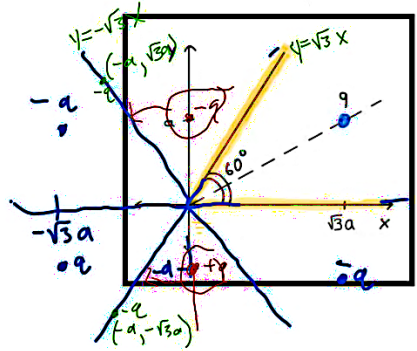

Fig. 17. The correct number of image charges obtained by G2 in Q4

Table 7. Summary of sensemaking of G2 in case C

| Aspect | Summary of G2's Sensemaking in Case C |
|---|---|
| Initial guess about image charges based on prior problem (local coherence and consistency across problems) | Started by guessing that two image charges would be required, likely based on prior experience (with Q2). |
| Need for diagrammatic representation | Entered Pictorial Analysis game; started adding axes and image charges on figure provided. |
| Nested epistemic games | Entered Transliteration to Mathematics game within Pictorial Analysis game to compare two problems (Q2 and Q3) in which charge was present between two conducting semi-infinite grounded planes. |
| Reuse of knowledge resources about image charge magnitude being q/2 (local coherence and consistency across problems) | Activated previously used knowledge (two image charges of magnitude q/2) from Q2, even though it was incorrect for Q3. |
| Iterative refinement of figures | Drew multiple diagrams to better understand the geometry, symmetry, and placement of image charges; each iteration improved spatial reasoning. |
| Checked boundary conditions | Chosen image charge configurations couldn't satisfy boundary conditions. |
| Scaffolding helped recognize symmetry | Scaffolding in Q4 helped in recognizing the symmetry of the problem; eventually correctly found five image charges with minor issues with position of charges. |
| Expert-like evolution | Reasoning evolved over time: started with incorrect number, position, and magnitude of image charges, but ended with correct number, correct magnitude, and almost correct positions. |

**Discussion of G3 for case C:** The sensemaking of G3 for case C is discussed in the Appendix. Summary of the sensemaking of G3 in case C is discussed in Table 8.

Table 8. Summary of sensemaking of G3 in case C

| Aspect | Summary of G3's Sensemaking in Case C |
|---|---|
| Use of diagrammatic representation | Entered Pictorial Analysis game; drew axes and charges on figure provided to visualize regions and symmetry; shaded regions to distinguish region-of-interest from excluded regions. |
| Nested epistemic games | Entered Transliteration to Mathematics game within Pictorial Analysis game; used knowledge from previous problems and reasoning about image charges while drawing diagrams and considering symmetry. |
| Consistent sensemaking about charge separation on conductors | Misinterpreted "grounded" as neutral/zero net charge instead of zero potential; assumed original charge induces charge separation on both sides of grounded conductor. Used knowledge from previous problem (Q2). |
| Consistent in searching for logic about number of image charges across problems | Initially related number of image charges to number of excluded regions; revised reasoning based on symmetry to eventually identify five image charges. |
| Role of scaffolding in recognizing symmetry | Scaffolding in Q4 helped in recognizing spatial symmetry of the space. |
| Conflict in deciding between three vs. five image charges | Even after recognizing symmetry, conflicted in deciding if there should be three or five image charges; easily discarded option of four or six. |
| Iterative refinement of figures | Redrew diagrams multiple times, adjusting placement of image charges to match perceived symmetry. |
| Correct image charge configuration despite inconsistent reasoning | Found correct number of image charges although idea of charge separation on the grounded conducting plane was not consistent. |
| Consistent struggle with distance interpretation in potential | Calculated distances from origin to point charges rather than from a general point to point charges; expression for potential was therefore not valid for a general point. |

**Discussion of G1 for case C:** We now discuss the sensemaking of G1 for the unscaffolded and then scaffolded versions (Q3 and Q4) and show how their approach evolved during this process. As discussed earlier in all other cases of sensemaking regarding MoI problems, G1 also started solving Q3 by adding details in Fig. 2 provided to the students and entered the Pictorial Analysis game. They started by marking the angle and the potential on the conducting planes as shown in Fig. 18.

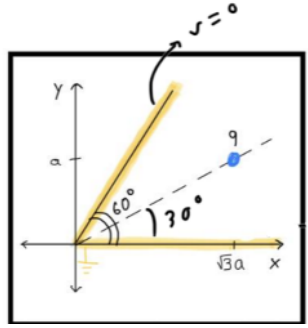

Fig. 18. G1 labeled boundary condition on one surface in Fig. 2 provided with Q3

To explore the geometry of this problem, which is a critical part of problem solving, especially for this complex problem, they decided to draw a new picture, shown in Fig. 19 (a), from scratch. They started by drawing the axes, grounded conducting planes, and the original charge, saying, *'...So if I add one charge* ['-q'] *same distance over here* [in the fourth quadrant as shown in Fig. 19 (a)], *so from here it's gonna be '−a'...this as ' $\sqrt{3}a$ '* [placed a charge '-q' at $x = \sqrt{3}a, y = -a$]. *And then also...this side* [labeled the distance $y = a$ for the original charge] *is 'a'...so this* [labeled the distance on the plane at 60°] *also has to be '$\sqrt{3}a$' so that these are similar triangles* [pointing to the interior region between the two planes] *...If I continue this line* [from plane at 60° to +y axis] *...this* [pointing to the distance from plane at 60° to +y axis] *is gonna be 'a'...'*.

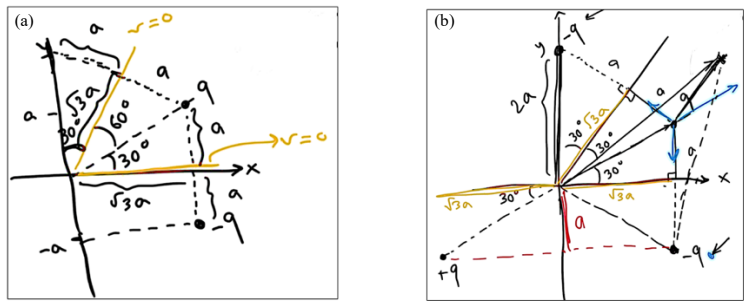

Fig. 19. (a) G1's drawing to find the positions of the image charges using geometry for Q3(left), (b) picture redrawn by G1 to find three image charges for Q3 (right)

Then G1 decided to draw the picture again saying, *'I don't know if I'm visualizing this right or not. Let me look at it over here again.'* Then they started by drawing the axes and the grounded conducting planes in Fig. 19 (b) and placed the original charge at the appropriate location in the picture. They labeled the distances in Fig. 19 (b), the same as those obtained in Fig. 19 (a) saying, *'...If I want to put the mirror image of this* [original charge as shown in Fig. 19 (b)] *on the other side...I'm gonna have to put my second image charge over here* [placed an image charge '-q' charge at $x = 0, y = 2a$] *and I have the other one to be over here* [placed another image charge '-q' at $x = \sqrt{3}a, y = -a$]. *Now the question is whether this is going to satisfy the boundary conditions of the original problem or not?'*. At this point, G1 entered the Transliteration to Mathematics game as they tried to relate the problem to another similar problem involving force vectors and said, *'...I* [can] *look at this problem like one of those problems...*[where] *we looked at the addition of all of the fields or the forces on a point charge...From this point charge* [pointing to the image charge in the fourth quadrant] *to this one* [pointing to the original charge], *we're gonna have this vector* [drew a vector pointing from original charge to the image charge as shown in Fig. 19 (b)] *and from this one* [pointing to the image charge on y axis] *to this one* [pointing to the original charge], *we're gonna have this* [drew a vector pointing from original charge to the image charge]. At this point, G1 thought that there should be another vector pointing opposite to the resultant vector of these two vectors such that the net force is zero, saying, *'...So I'm gonna end up needing something that cancels out these vectors like this* [drew a third vector at the original charge pointing away from the origin]. *I want to have the vector going that way* [points to the direction they drew on the picture away from the origin...*I need to have '+q''*. Then, G1 compared this situation with the problem given in Q2 and thought that this problem will also have three image charges similar to Q2 saying, *'...sort of similar to the...example we had, a point charge here* [drew the picture for Q2 not shown here], *and I added one* [image charge] *here* [in the second quadrant], *one here* [in the third quadrant], *and then one here* [in the fourth quadrant]'. G1 came back to the original problem and placed the third image charge in the third quadrant at x = $-\sqrt{3}$a, y = $-$a. Then G1 wrote down an expression for the potential using the original charge and the three image charges, which was not correct because it had only three image charges instead of five (not shown here). The expression for potential with three image charges and the original charge was mathematically correct, but could not satisfy the boundary conditions.

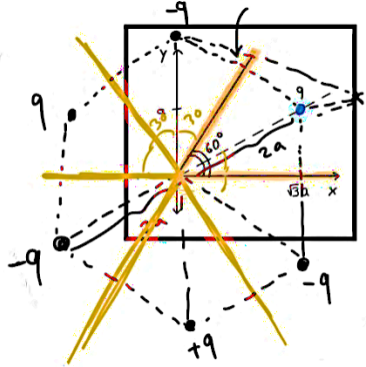

Fig. 20. G1's drawing shows five image charges for Q4

When G1 was given this same problem in the scaffolded version, their approach showed further improvement. Fig. 20 shows the final picture drawn by G1 after solving the scaffolded problem. At the beginning of the problem-solving process, G1 read Q4 which was given as part of the scaffolded version. After reading the options for the number of image charges possible they said, '...Well, I used three [image charges earlier in the unscaffolded version] but let's think, maybe I was wrong...I might have wanted to use more...'. They realized that they should take a look at their vector diagram for the forces on the original charge from the image charges and they started drawing it as shown in Fig. 21 saying, '...Oh, maybe what I didn't take into account when I was showing the vectors, I had a '-q' here [referring to the image charge placed in the fourth quadrant] and then I added another '-q' over here [referring to the image charge placed on +y axis in Fig. 20]...For the distance 2a I'm going to take 'A' [they labeled two vectors in Fig. 21]...and then for this one [third image charge added in the third quadrant] that I added, I put it over here as 'q', this would be 4a [referring to the distance between the image charge in the third quadrant and the original charge], this would be 2A [pointing to the length of the third vector in Fig. 20]...I wanted them [vectors] to...cancel each other out. I have this is 30° [labeled the angle between the x-axis and 2A] and this is 60° [labeled the angle between 2A and the y-axis]...This 2A is going to be 2Acos (30°) [pointing to projection vector along x-axis]...This one is going to be 2Asin (30°)...they wouldn't cancel out.' At this point, G1 realized that their earlier approach regarding vectors was incorrect. They drew another picture of the vectors, saying, '...I wanted them all to look like this actually, instead [drew the picture as shown in Fig. 22]. So this way I would have been able to use the method of images and say that, okay, I have the same boundary conditions'. G1 thought that the correct force diagram would have helped in satisfying the boundary conditions (which is not a correct approach). They also thought that there were multiple image charge configurations that would be appropriate for this problem, saying changing the magnitude of the charges could be another way to get the correct vectors, 'I [could] have possibly turned this [pointing to the image charge in the third quadrant] into $\frac{q}{2}$...'. But G1 did not proceed with this idea, saying, '...Instead, if I don't want to be dealing with how much I'm changing the [magnitude of] charges...I think I can look at this question as I have 60° [drew a line in the fourth quadrant making an angle 60° with the x-axis as shown in Fig. 20]. This [pointing to y-axis from the second grounded conducting plane] is at 30°. So, this [drew a line in the second quadrant making an angle 30° from y-axis] also needs to be at 30°...'. With this thought, G1 drew all of the symmetry lines as shown in Fig. 20 and realized that the figure looks more symmetric now saying, '...So based on the symmetry that I want to preserve, I would have added -q here [image charge on +y axis] and then I would have again added 'q' over here [image charge in the second quadrant] and then I would have added this [image charge '-q' in the third quadrant]...this has to be positive [the image charge '+q' on -y axis], so that this one would be negative [the image charge '-q' in the fourth quadrant].' At this point, G1 realized that this problem can be solved with five image charges saying, I think in that case I would need to add five image charges. But there might also be a way that I could control for the amount of the mirror charge that I am putting or maybe the distance and then only use 3 image charges. So, I'm also doubting on this one [whether this problem can be solved with 5 image charges or whether 5 or 3 image charges could both be fine if the distances were chosen appropriately in each case]. It's definitely not 4 or 6 [image charges]'. Summary of sensemaking of G1 in case C is discussed in Table 9.

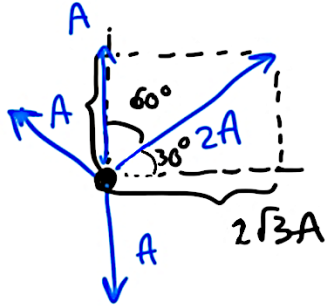

Fig. 21. Force vectors on the original charge (image charges introduced to make the net force zero) drawn by G1 for Q4

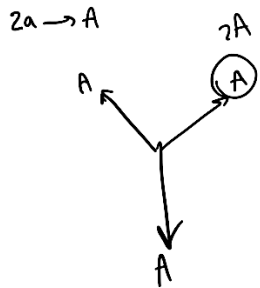

Fig. 22. Force vectors on the original charge (image charges introduced to make the net force zero) drawn by G1 for Q4

Table 9. Summary of sensemaking of G1 in case C.

| Aspect | Summary of G1's Sensemaking in Case C |
|---|---|
| Use of Diagrammatic representation | Entered Pictorial Analysis game; added axes and image charges on figure provided. |
| Sensemaking involving Geometric exploration | Used knowledge of similar triangles to explore geometry; drew axes, grounded conducting planes, and original charge. |
| Iterative refinement of figures | Drew multiple pictures iteratively to clarify geometry and placement of charges; each drawing helped refine spatial understanding. |
| Initial guess about image charges based on prior problem-solving (local coherence and consistency across problems) | Started with three image charges based on prior experience (Q2); tried to justify their choice. |
| Sensemaking about boundary conditions showing local coherence | Knew boundary conditions must be satisfied to justify locations and magnitudes of image charges. |
| Consistent sensemaking about idea of "balancing" | Carried along idea of "balancing forces/charges" from Q1 to Q3/Q4; initially thought forces being "balanced" implied boundary conditions were satisfied. |
| Consideration of making image charge magnitude q/2 to satisfy boundary condition | Considered adjusting magnitude of image charges to q/2 (similar to G2 in Q2/Q3) but did not proceed further. |
| Nested epistemic games | Entered Transliteration to Mathematics game to relate problem to similar past problems (involving vectors, forces); nested this within Pictorial Analysis game. |

| Conflict between three vs. five image charges | Encountered a conflict regarding whether there should be three or five image charges based on symmetry. |
|---|---|
| Sensemaking involving configurational symmetry | Multiple attempts activated idea of symmetry; drew six-fold symmetry lines to guide placement, magnitude, and sign of image charges. |
| Expert-like evolution | Reasoning evolved from inconsistent justification involving three charges to correctly identifying number, magnitude, sign, and approximate positions of five image charges through symmetry considerations and iterative diagrams. |

**Case C summary:** In the three cases discussed here, we observed that grad students sometimes carried along concepts and activated the same knowledge resources across different problems, which might not always be correct. As discussed earlier, G2 carried along the (incorrect) concept from the previous problem (Q2), as they used q/2 in the first attempt of a new problem (Q3). G3 also incorrectly used the concept of charge separation and interpreted the grounding of the conductor incorrectly in Q2 and Q3. Also, G1 kept using the idea of balancing charges/forces in different contexts. We also observed that G3 could not interpret the region of interest correctly at the beginning but realized soon that the plane should be divided into six parts instead of four to obtain the correct symmetry. We also observed that when students attempted a problem multiple times and drew diagrams for the same problem multiple times, there was an improvement in each attempt, and they were getting closer to solving the problem correctly. G1 was provided with the unscaffolded version of the problem first and then got the scaffolded version immediately after that. Initially, instead of looking at the potential at the boundary, G1 incorrectly focused on obtaining the vector sum of all the forces on the original point charge to be zero. They thought that when the net force from the image charges on the original point charge sums up to zero, it would automatically satisfy the boundary conditions, but they did not check the potential at the boundaries to verify this. We also observed that when students were asked in the scaffolded version of the problem to choose the correct number of image charges and were provided with the options of three, four, five, and six, this small scaffolding of providing a separate problem in which they had to contemplate the options helped them to think about the symmetry as they had to find reasons to discard four and six image charge options provided. Thus, the question with options about the number of image charges may have helped and prompted some students to think about the correct choice involving five image charges. Thus, sometimes intermediate questions providing small scaffolding can help in activating appropriate knowledge resources to solve the problem correctly. Key sensemaking patterns and primary epistemic games played by G1, G3, and G2 in case C are discussed in Table 10.

Table 10. Key sensemaking patterns and primary epistemic games played by G2, G3, and G1 in case C

| Student | Key Sensemaking Patterns | Primary Epistemic games (nested) |
|---|---|---|
| G2 | Reuse of knowledge resources about image charge magnitude being q/2 (local coherence and consistency across problems).<br><br>Checked boundary conditions. | Q3<br>Pictorial Analysis<br>   Transliteration to Mathematics<br>   Mapping Meaning to Mathematics<br>Q4<br>Pictorial Analysis<br>   Mapping Meaning to Mathematics |
| G3 | Consistent sensemaking about charge separation on conductors.<br><br>Consistent in searching for logic about number of image charges across problems.<br><br>Correct image charge configuration despite inconsistent reasoning. | Q4<br>Pictorial Analysis<br>   Transliteration to Mathematics<br>   Mapping Meaning to Mathematics |
| G1 | Sensemaking involving geometric exploration. | Q3<br>Pictorial Analysis |

| | Sensemaking about boundary conditions showing local coherence. | Transliteration to Mathematics |
| | | Q4 |
| | Consistent sensemaking about the idea of balancing. | Pictorial Analysis |
| | Consideration of making image charge magnitude q/2 to satisfy boundary condition. | |
| | Sensemaking involving configurational symmetry. | |

## IV. BROADER DISCUSSION

In this research, we focused on graduate students' sensemaking using the epistemic game framework while they solved problems involving the MoI. Understanding how grad students think about these types of problems and what could be an appropriate level of scaffolding can enable instructors and researchers to help students learn physics better. Very few prior studies have focused on the sensemaking of advanced physics students, and to our knowledge only Ref. [29, 71] used the epistemic game framework to investigate the sensemaking of physics graduate students in the upper-level electricity and magnetism course. We focused on three grad students' sensemaking episodes that stood out to the researchers as valuable for their pedagogical implications. We adapted Tuminaro and Redish's [66] epistemic game framework, which was originally developed for understanding the sensemaking of introductory physics students. In particular, we observed graduate students playing epistemic games, which had structural components such as the moves of the game, similar to the ones that Tuminaro and Redish described in the introductory physics context. However, the ontological components of some of the games that the grad students played in our investigation are different. For example, in some of the epistemic games, introductory students activated intuitive knowledge from their everyday life exclusively [66], whereas grad students in this study played epistemic games with similar moves that involved the activation of intuitive knowledge that was learned in formal contexts (not exclusively from everyday life situations). This section includes discussion about findings for both physics educators, particularly focusing on the method of images or related concepts, and for a broader audience. Table 11 shows the prominent sensemaking patterns observed across three cases and G1, G2, and G3. In the following discussion, we relate our findings to the different research questions.

Table 11. Prominent sensemaking patterns observed across cases A, B, and C and students G1, G2, and G3

| Case | Students | Prominent sensemaking patterns |
| --- | --- | --- |
| A | G1 | Use of diagrammatic representation. |
| | | Nested epistemic games. |
| | | Mixing of concepts. |
| | | Effects of nudging / interviewer's prompts/scaffolding. |
| | | Struggle with distance interpretation in potential. |
| | | Persistence of "balancing" idea. |
| B | G1, G2, G3 | Use of diagrammatic representation. |
| | | Nested epistemic games. |
| | | Iterative refinement of figures. |
| | | Struggle with distance interpretation in potential. |
| | | Local coherence during sensemaking. |

|   |   | Persistence of certain ideas across different problems |
|---|---|---|
| C | G1, G2, G3 | Initial guess about image charges based on prior problem solved (local coherence and consistency across problems). |
|   |   | Use of diagrammatic representation. |
|   |   | Nested epistemic games. |
|   |   | Iterative refinement of figures. |
|   |   | Scaffolding helped recognize symmetry. |
|   |   | Expert-like evolution. |
|   |   | Local coherence and consistency across different problems. |
|   |   | Conflict in deciding three or five image charges. |
|   |   | Consistent struggle with distance interpretation for potential. |

### A. Sensemaking about the Method of Images

**Interference and mixing of related concepts (RQ1):** Regarding RQ1, grad students' sensemaking shows that they activated knowledge resources involving the mixing up of different related concepts. For example, in case A, G1 was confused between the total charge (e.g., total charge within a Gaussian surface is relevant for the integral form of Gauss's law) and the local charge density (e.g., relevant for the differential form of Gauss's law from which the Poisson equation follows when one relates the electric field to the potential). G1 also confused two methods for finding the potential often taught in advanced electricity and magnetism courses, and thought that the MoI involves selecting the image charges to turn Poisson's equation into Laplace's equation. This type of issue is common in introductory physics as sometimes students struggle to differentiate, e.g., between the force and momentum [72], the resistance and resistivity [73], or the conductor and insulator [74]. However, this type of confusion suggests that even for graduate students, related concepts learned within a very short interval can interfere with each other, and an amalgam of related concepts may get activated while solving problems.

**Use of reasoning primitives and using them across different problems (RQ1):** Tuminaro and Redish also discussed [66] how introductory students used different reasoning primitives, e.g., "balancing", "canceling", "more is more", etc., while playing different epistemic games. Related to RQ1, we observed that even in the case of grad students' sensemaking discussed here, we see evidence of the use of reasoning primitives. For example, in cases A and C, we observed that G1 used the idea of 'balancing' charges or forces across different MoI problems. In the first MoI problem they solved, G1 believed that their of placing the image charges was to "balance" the charges (i.e., to make the net charge including all regions zero) so that Poisson's equation becomes Laplace's equation. Then G1 contemplated (in the same problem) whether the charges should be 'balanced' in such a way that the net charge distributed above and below the origin on the y axis would be the same. Later, while doing sensemaking in the context of the MoI problem involving the sextant, G1 activated the knowledge resource pertaining to the "balancing" of forces on the original charge from the image charges. This reasoning primitive resource was activated several times while solving the sextant problem in unscaffolded and scaffolded forms and lingered until the end in the student's mind, even though the student recognized the hexagonal symmetry of the problem and solved the problem correctly, likely without complete understanding. In particular, the grad student had not completely ruled out that other approaches, such as balancing of forces, would also help to figure out the positions of all the image charges. Also, in cases B and C, we observed that grad student G3 activated the 'grounded as neutral' reasoning primitive (they still obtained the correct image charges because they understood the symmetry of the charge distribution, even though they incorrectly assumed throughout the sensemaking process that there was a charge separation on the grounded conductor). This example also shows that sometimes students may end up with the correct solution without using completely correct reasoning, e.g., by exploiting the symmetry correctly in this case.

**Activation of a resource not relevant to problems multiple times across different problems (RQ1):** Related to RQ1, we also observed that during grad students' sensemaking, if no scaffolding was provided, a particular resource not relevant in a particular context was activated multiple times across different MoI problems. For example, G2 checked the boundary conditions at the origin and at infinity in case B, which would work for Q1, but for Q2, which involves two planes, students must check that the boundary conditions are satisfied at other points on the surface (which G2 did not do). However, the boundary conditions being satisfied for the few points G2 checked gave them the incorrect perception of being correct about the image charges having a magnitude q/2, and they carried this incorrect concept of image charges being q/2 to other problems, e.g., as observed in case C for Q3. Similarly, in the preceding discussion about the reasoning primitives, G1 had a notion of balancing the charges or balancing forces acting on the original charge. We observe that even when G1 progressed from Q1 to Q3 in the problem-solving process, they continued to activate the resource of 'balancing' in the context of forces (in case C) from the balancing of charges (in case A). Thus, the balancing resource was readily activated in the MoI problems for them. They also incorrectly thought that satisfying the boundary conditions for a MoI problem is equivalent to the balancing of the forces on the original charge due to the image charges (even though only the boundary conditions are relevant for the MoI problems posed). Thus, we find that sometimes grad students activate knowledge resources multiple times across problems (sometimes in adapted forms, e.g., balancing of charges vs forces) that might not be consistent with the physics involved, even though they are useful in other situations. This suggests that providing adequate scaffolding support, e.g., via check points, after students have struggled with an issue and activated a resource in a context in which it is not applicable, can ensure that they do not continue to activate the same irrelevant resource across different problems.

**Struggle differentiating between aspects of problems that are unique or not unique (RQ1):** In relation to RQ1, our findings suggest that it would be valuable to help students differentiate between aspects of problems that are unique or not unique. For example, in case C, we observed that G1 was able to beautifully evolve through the problem-solving process and was able to explore the symmetry extremely well at the end in an expert like manner. However, G1 still continued to think that the problem could potentially be solved with three image charges instead of five. Their continued support for the idea that three image charges might also be appropriate if they had the correct magnitude and sign and were placed in appropriate places may have been reinforced by G1's earlier attempts at the problem. In particular, G1 was trying multiple approaches to put image charges while solving the problem in case A, as well as contemplating the idea that there may be multiple ways to place image charges to solve the problem in case C also. This type of continued struggle suggests that it may be useful to help students be able to differentiate between the aspects of a physics problem that are unique, e.g., in this case, the image charge configuration is unique (cannot involve multiple ways of placing the image charges) and the aspects that are not unique, e.g., physics problems can be solved using multiple methods even though some are generally more effective approaches than others.

**Common challenges in juggling different components of MoI problems (RQ2):** In relation to RQ2, we observed that while sensemaking, the common trend seen across grad students and across different cases was that students struggled in figuring out the correct number, position, magnitude, and sign of the image charges. For example, we find that when students were able to recognize the method correctly, they often struggled in figuring out the correct number of image charges. This is not an easy task, as students need to understand the symmetry of the problem to figure out the number of image charges correctly. Furthermore, even students who understood the importance of checking the boundary conditions to make sure they obtained the correct magnitude, sign, and position of image charges, struggled to ensure that the boundary conditions were satisfied appropriately in many cases.

**Struggle with distance interpretation in potential (RQ2):** Regarding RQ2, we also find that grad students often struggled to interpret the 'distance' in the expression for the potential, something that has been observed for introductory students in the context of electric field [75]. The struggle with the correct interpretation of 'distance' was commonly observed across multiple problems, across different students. These struggles suggest the importance of helping students understand, in physical terms where the potential is being evaluated. Also, this issue with distances, which was observed in the context of electric potential here, is likely to surface when calculating other physical quantities involving distances, such as electric field [75].

**Emphasis on symmetry (RQ2):** In relation to RQ2, our findings suggest that some students were able to evolve in their sensemaking process and progressively harnessed the symmetry of the charge distribution involved. However, it is important to continue to emphasize the symmetry inherent in physics problems to help grad students learn how to exploit the symmetry appropriately. For example, while symmetry considerations are important in general in MoI problems, in the most complex problem (case C), the appropriate exploration of symmetry plays a central role in being able to solve the problem, and grad students struggled with it.

### B. Broader Discussion: Sensemaking, Epistemic Games, and Knowledge Activation

**Activation of prior knowledge, local coherence in sensemaking, and consistency across different problems (RQ1, RQ2):** Related to RQ1 and RQ2, we observed that students' sensemaking shows that they frequently activated valuable resources, drawing on their prior knowledge to navigate different problems. While knowledge activated is often constructive and helpful, there were also repeated activations of resources that were inconsistent with the problem contexts but appeared reasonable to

students in the sensemaking observed. In particular, we observed that students often demonstrated local coherence, looking for internal consistency locally in their approach during sensemaking. However, their reasoning often lacked global coherence, leading to solutions that do not fully align with relevant physics concepts. Understanding these patterns of sensemaking can help instructors provide targeted support at appropriate moments.

**Improved problem-solving approaches with successive attempts and multiple drawings (RQ2):** Regarding RQ2, grad students' problem-solving approaches improved with successive attempts. In particular, we find that the grad students often evolved and became more expert-like in their problem solving in successive attempts. For example, in both cases B and C, grad students often attempted to refine their approaches while making sense, e.g., by drawing many more diagrams than asked for explicitly in the problem by re-drawing, them from scratch. While students did not have any dedicated course work or tutorial scaffolding in drawing 3D diagrams or diagrams from multiple perspectives to help in placing the image charges (except there were prompts that explicitly asked them to draw figures), students used more 3D looking diagrams with the grounded conducting planes to visualize and figure out where to put the image charges. Once they had developed intuition about where to put the image charges, they often found it more convenient to draw 2D configurations of all the point charges to find the potential. The thoughtful sensemaking of G3 is epitomized in their reflection, '*I didn't really draw this nicely…I'm gonna try drawing this in 2D, maybe it will help me think about things differently… maybe I'll redraw this a couple of times and that'll help me think about it more clearly.*' We observed that with successive attempts at drawing these diagrams, students generally got better, and the improved diagrams helped them reason in a way that their solution became closer to an expert-like solution (even if there were still some lingering issues). Grad students' sensemaking suggests that it is important for instructors to continue to emphasize the usefulness of multiple representations in solving physics problems effectively. The MoI problems discussed here rely on the effective use of diagrammatic representation. If students struggle to use the representations effectively, it can become a hindrance in problem-solving. Instructors can focus on encouraging and incentivizing students to make multiple attempts at a problem, e.g., by re-drawing diagrams for a problem and thinking about it again, as it can help improve their problem-solving.

**Nested structure of the epistemic games (RQ3):** Regarding RQ3, we observed that advanced students' sensemaking using different epistemic games often had a nested structure as grad students switched between different games or started playing a new game within the moves of another game that they came back to later. In a study focused on sensemaking of upper-level students using four epistemic resources (calculation, physical mapping, invoking authority, and mathematical consistency), Bing and Redish [38] discussed the nesting of subframes within a larger coherency valuing frame similar to the nested structure of the epistemic games we found in this investigation in the context of graduate students' sensemaking. On the other hand, since in Tuminaro and Redish's paper describing the epistemic games played by introductory students [66] short episodes are described, it is unclear whether, in extended episodes, even introductory physics students' sensemaking would show evidence of nested epistemic games. Even if that is the case, it is likely that the nested structure of the epistemic games is more complex in the case of advanced students' sensemaking as discussed in this investigation due to the more complex nature of the problems.

**Impact of nudging and scaffolding while sensemaking (RQ4):** While discussing conceptual blending, Hu and Rebello [6] observed how interviewers' nudges/questions can cause students to significantly change their problem-solving approaches. Regarding RQ4, we observed large effects of nudges in some cases. For example, in case A, when grad student G1, who had ended the game without finding the desired expression for potential, was asked by the interviewer to think about the expression for the potential, G1's framing of the problem changed. They resumed the problem-solving process and wrote down an expression for the potential activating their knowledge resources related to the potential for a point charge and the superposition principle (which they had not activated earlier) instead of claiming that they would need to solve Laplace's equation as they did earlier. Similarly, we observed that when a small amount of scaffolding support was provided to G1, G2, or G3 as discussed in case C, it was often helpful as it steered them in the right direction. For example, in case C, the scaffolding provided by an additional question providing possible options for the correct answer helped students rethink their strategies and figure out the correct configuration. Nudging and guidance could help students activate relevant knowledge resources at appropriate points. The examples of students discussed here demonstrate how targeted prompts can redirect sensemaking effectively. Thus, if we can identify those critical points at which a certain level of scaffolding or nudging is most useful, we can better provide support to help students learn and do more expert-like sensemaking.

## V. CONCLUSIONS AND COMPARISON WITH SENSEMAKING IN LAPLACE EQUATION CONTEXT

Overall, graduate students' sensemaking discussed here shows evidence of their evolving expertise and local consistency and coherence, even when it lacks global coherence. All the instances show that observing student sensemaking is valuable to understand how students reason about concepts while solving physics problems and what obstacles can hinder their path to successful problem solving. Understanding the dynamics of student sensemaking and how they tackle situations in which they are stuck or how they contemplate multiple strategies to move forward can help instructors and researchers to focus on effective instructional approaches with an appropriate level of scaffolding. Without observing student sensemaking, there may be misalignment between what instructors think would support them and the support that would actually help students. Observing sensemaking helps us to remove the veil from those hidden parts which remain unobserved, and this research shows that these issues are not only important in introductory physics, as discussed in prior studies but also in advanced physics.

Some features of the MoI sensemaking research findings discussed here, such as the nesting of the epistemic games or mixing of methods, are similar to sensemaking in the context of LE [29]. However, this study in the context of MoI sensemaking reveals several novel findings that are different from the earlier investigation in the context of LE [29]. For example, unlike Ref. [29], in the investigation presented here, not only do we make in-depth comparisons between multiple students, but also in-depth comparisons are made between related but distinct problems for individual students. Some of the novel findings in the context of MoI sensemaking are as follows: This study investigated the impact of small nudging and scaffolding that showed significant improvements in the way grad students approached problem-solving. Another novel finding is that students showed persistence in activating different knowledge resources multiple times across different problems, even if those resources were not applicable. The study showed consistent use of reasoning primitives across different problems such as a "balancing" primitive multiple times while sensemaking. Findings also reveal the common challenges grad students experience while juggling different components of MoI problems involving the correct position and magnitude of image charges, along with their correct number by making sure all the boundary conditions are satisfied and not just some of them. Another novel finding is that grad student problem-solving approaches significantly improved with successive attempts and multiple drawings for the same problem and across problems. Often, students initially drew more 3D figures with conducting planes to visualize the problem better; later, they also drew diagrams from different viewing angles, which helped improve their physical understanding and comprehension of the symmetry of the underlying problem. Grad students struggled to differentiate between aspects of problems that are unique and not unique. For example, they thought that in a given problem involving the MoI, the unique electric potential can be obtained by either three or five image charges if the magnitude and position of image charges are chosen appropriately, which is not possible according to the uniqueness theorem.

## VI. INSTRUCTIONAL IMPLICATIONS, LIMITATIONS, AND FUTURE DIRECTIONS

The findings of this research highlight several important implications based on the sensemaking of advanced students. First, scaffolding strategies play a crucial role in helping students navigate complex geometries. Instructors can provide explicit scaffolds, such as prompts that focus attention on symmetry and verification, and design problem sequences that gradually increase in geometric complexity with structured intermediate steps. Second, supporting 3D visualization is essential for developing a deep understanding of the spatial aspects of physics problems. Students should be taught multiple visualization strategies, including 3D perspectives, 2D projections, and symmetry-based reasoning. Physics instructors should emphasize the importance of spatial accuracy in diagrams and encourage iterative refinement as a normal part of problem-solving, while also providing feedback on diagram quality and spatial reasoning, not just on final answers. Third, addressing transfer and overgeneralization can help students discern when specific intuitions or methods are applicable. Explicit comparison of problems before solving can help students analyze structural differences and identify which features are general versus context dependent. Encouraging metacognitive reflection about when and why a particular method transfers can prevent the unproductive reuse of prior reasoning. Fourth, addressing common conceptual difficulties is key. Physics instructors should make clear distinctions between related but distinct ideas. Emphasizing how geometric symmetry relates to the number of image charges and illustrating how boundary conditions relate to determining the signs of image charges, can support conceptual clarity. Finally, supporting sensemaking as an iterative and exploratory process is vital. Students should be given time and space to explore, test, and revise their reasoning without the expectation of immediate correctness. Think-aloud discussions and peer collaboration can help externalize reasoning, while physics instructor "nudges", such as asking whether boundary conditions have been checked everywhere as appropriate can gently guide students toward deeper understanding. Recognizing that struggle and revision in the problem-solving process are productive elements of learning can foster a more supportive and authentic environment for sensemaking and developing expertise in solving physics problems.

One limitation of this study is that it involves in-depth interviews with a small number of graduate students from a large public research university in the United States. Also, although graduate students' sensemaking about upper-level electricity and magnetism problems involving MoI is likely to be similar to that of upper-level undergraduates who have learned these concepts, the focus on grad students' sensemaking may be viewed as a limitation regarding the extent to which these results are directly generalizable to upper-level undergraduates. However, the graduate-level physics course in electricity and magnetism also includes the types of problems discussed here, so grad students' sensemaking is insightful for understanding how they activate knowledge resources, how they evolve with successive attempts at the same or different problems related to MoI, and how scaffolding impacts their evolution. Future studies should investigate sense-making of advanced students, including graduate students and upper-level undergraduates, across different physics concepts at different types of institutions internationally. Since this study was conducted at a single large public university, other studies at different types of institutions may show that advanced students' sensemaking in a given physics context, e.g., while solving MoI problems discussed here, varies depending on their prior preparation and institutional context.

## ACKNOWLEDGMENTS

We dedicate this paper to the memory of Joe Redish. We thank the grad students who participated in this study. We thank Dr. Robert P. Devaty for his constructive feedback on the manuscript.

## APPENDIX

**Discussion of G3 for case C:** We now discuss the sensemaking of another grad student, G3. This student did not spend a significant amount of time in sensemaking in the unscaffolded version, and therefore, we will not discuss the unscaffolded version and only focus on their sensemaking in the scaffolded version of the problem. In the scaffolded version, they had to find the correct number of image charges first, as given in Q4, and then they were asked to write the expression for the potential for the same problem (that part of the problem is not shown with Q4 in this paper but is similar to Q3). G3 decided to add details to Fig. 2, which was also given in the scaffolded version and entered the Pictorial Analysis game. They started by shading the sextant region of Fig. 2. We note that G3 incorrectly considered half of the area as the sextant, as shown as the shaded area in Fig. 23, whereas it should be the total region between the two grounded conducting planes.

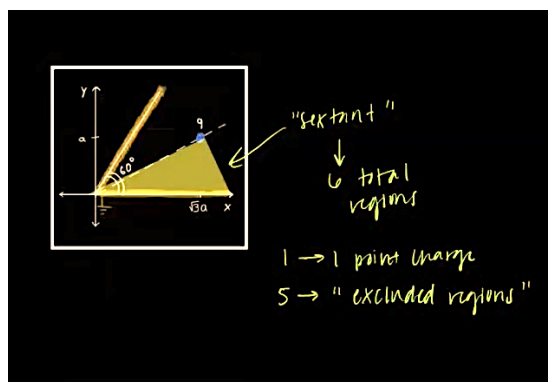

Fig. 23. Picture drawn by G3 where they incorrectly shaded half of the region of interest as sextant for Q4

They thought that each sextant is a different excluded region and there is some relation between the number of those excluded regions and the number of image charges, saying, *'...here* [pointing to the shaded region in Fig. 23] *is what they call a sextant, meaning there are six total regions. One of them has one point charge right now, and we have five other excluded regions...I'm going off that logic where it's* [number of image charges] *not necessarily proportional to the* [original] *point charge itself, it's more...proportional to these other* [excluded] *regions...'*. Then, within the move of the Pictorial Analysis game, they entered the Transliteration to Mathematics game as they started comparing the given problem with another simpler problem based on the MoI, in which a point charge is present in front of an infinitely long grounded conducting plane placed in the $xy$ plane as given in Fig. 1. Then they drew the picture of the comparison problem shown in Fig. 24 and appeared to be going back to the Pictorial Analysis game with the goal of finding a relationship between the number of image charges and the number of excluded regions (which could be multiple excluded regions according to G3) in the problem.

G3 said, *'...if I'm looking back at the simple case of the plane conductor and the zy plane* [drew $zy$ axes as shown in Fig. 24], *and I think in that problem, we had one charge here on the z axis* [placed a charge 'q' on the positive $z$ axis], *and then we were able to find that there's another charge on this side like using method images* [placed an image charge '-q' on −z axis]. G3 thought that the original charge 'q' leads to a charge separation on the surface of the infinite grounded conductor [not explicitly shown in Fig. 24] due to which a negative charge is induced on the top surface of the infinite grounded conductor facing the charge and a positive charge is induced on the other side of the surface. G3's sensemaking focused on the positive charge induced on the other side of the surface leading to the negative image charge [which they drew as an image charge in Fig. 24] in the excluded region saying, *'And if we were looking at the charge separation, then in that case, here was one region* [shaded the region above the $xy$ plane] *and this other side was the excluded region* [region below the $xy$ plane]*...'*. G3 focused on finding the relationship between the number of image charges and the number of excluded regions saying, *'...so if I'm using that logic, there should be one image charge for every excluded section* [region]*, I'll say…'*. They started focusing on each quadrant saying, *'...one image charge per excluded region. But for this simple case* [based on Fig. 24]*, we were looking at it in terms of quadrants like the first quadrant, the second one, the third and…the fourth. It* [number of image charges] *was basically just based on the sides of the conductor, and like the charge separation* [induced charge on both sides of the conducting plane as they misinterpreted the grounding]*…'*.

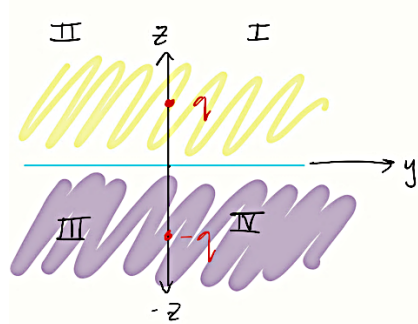

Fig. 24. Picture drawn by G3 while solving Q4 drawing analogy between Q4 and a simple MoI problem with one point charge on the z axis

G3 then went back to the original problem with the sextant given in Q4 and started drawing the induced charges on the surfaces, axes and the image charges in different sextants as shown in Fig. 25 saying, *'...So, we have this one* [highlighted the plane at 60°] *and this one* [highlighted the xz plane as shown in Fig. 25]*…if these are grounded…okay well, let me see if I can do this based on the charge separations. So, these* [both the conducting planes] *are supposed to both be grounded. So they're going to be neutral or have a neutral charge* [referring to the charge separation]*…well, we're gonna have negative charges, at least on this side* [drew negative charges on the inner side of the plane at 60° and also drew positive charges on the other side of the plane at 60°]*…there should be negative point charges in the opposite corners* [placed '-q' image charge below the positive $x$ axis and another '-q' charge on the left side of the positive y axis] *and then I know that these are gonna continue because they're intersecting* [extended the two conducting planes given in the problem Q4 by drawing planes passing through the origin along the negative x axis and at 60° angle to the negative x axis]. *So there's gonna be some other separation of charges on this side* [drew positive charges on the upper side of negative $xz$ plane and negative charges on the other side of the plane]*…That means there's gonna be a positive charge on this* [drew positive and negative charges on both sides of the extended side of the plane at 60 ° ]*…And why do I only get three charges? Maybe because the sextant is only supposed to refer to the position of the…point charge, and it's not really like a defined region.'* G3 thought that they got three image charges even when they found five excluded regions which contradicted with their initial assumption and made them think that there might not be any relationship between the number of excluded regions and number of image charges which they were earlier thinking about. G3 then realized that the region of interest should be the whole region between the two grounded conducting planes as shown by the shaded region in Fig. 25, and said, *'…this is actually the region of interest right here* [shaded the region between the two conducting planes including the blue and green regions in Fig. 25] *and the excluded regions are the other corners* [referring to the remaining regions]*, which I'll say is over here* [shaded the region between the grounded conducting plane at 60° and the $-x$ axis as shown in Fig. 25 with red color]*, this is one over here* [shaded the region between $+x$ axis and extended side of the grounded conducting plane at 60° in red color]*, and then the last one is the opposite* [shaded the region opposite to the region of interest in red color]*. Because those are the only other regions that are formed by this conductor. This is the same logic here* [number of excluded regions equals number of image charges]*, but it's not necessarily correct to the orientation of the conductors. So, the sextant* [is] *supposed to refer to something else. They're actually still quadrants formed for four total regions and three of them are referred to as the excluded regions. Okay, so in that sense, I will choose three image charges.'* Thus, student G3 thought that there was one region of interest where the original point charge was (colored green and blue in Fig. 25) along with three excluded regions with one image charge each (all three colored red in Fig. 25) and all these four regions can be thought of as "quadrants"(although those were not equal according to G3's drawing). G3 wondered what the relevance of the word sextant was in this context since they still ended up with four regions (region of interest and three excluded regions) similar to how they had reasoned about problem Q2 earlier.

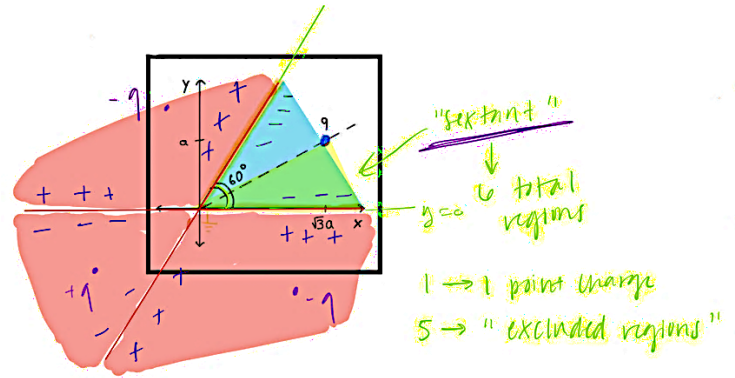

Fig. 25. The picture shows three image charges drawn by G3 for Q4

G3 then wrote down an expression for the potential using the original charge and the three image charges they obtained but that expression did not appear to be their desired expression for the potential that would satisfy the boundary conditions. Therefore, they decided to redraw the picture as shown in Fig. 26 saying, '…*Let me think if I…need…5 [image charges] over 3 [image charges]. I don't think I understand…how I would get 5 [image charges] though if that is the case. But then how does that really change my answers from what I have now?... I'm gonna redraw this because I think that's what's so…confusing. I can't figure out if it's three or five now* [started drawing the picture shown in Fig. 26]'. G3 started by drawing the axes and conducting planes extending across the origin and placed the original charge at the given position saying, '…*So if I follow the lines to make up these 6 sextants*…[this] *clarifies the sextant that refers to 6* [recall that G3 earlier thought that this problem Q4 also has four relevant "quadrants" similar to Q2 and they were unsure about how the word sextant was relevant for Q4 based upon their reasoning]*…because for the single one* [referring to the problem of a point charge placed in front of a grounded conductor]*, I didn't really have to do anything. I just knew that…the excluded region was just the other side. But for the intersection of two* [planes]*, what is the other side that we're talking about? Maybe I drew them in a way that doesn't represent the six sections, maybe mirroring them* [the charges] *will help.*' G3 placed five image charges on the symmetric axes drawn as shown in Fig. 26 and drew the induced charges on both sides of the axes saying, '*Well, this one is more symmetric, technically.*' They realized that they were dividing the regions into quadrants whereas it should be divided into sextants saying, '*…I think I was trying to fit these into quadrants but the quadrants aren't the same size* [as shown in Fig. 26]. *Intuitively, I wanted to say five* [image charges] *but then I thought that the definition of the sextant was kind of arbitrary. I guess, drawing it out into something more symmetric…makes more sense* [and clarifies] *that there* [should] *be 5 image charges.*' At this point, G3 entered Mapping Meaning to Mathematics and wrote down the expression for the potential using the original charge and the five image charges. They assumed that the distances of all the charges from the origin are the same and can be represented by a single variable $'r'$, which is '2a' for the original charge. Thus, they calculated the potential by writing $V = -\frac{kq}{r} + \frac{kq}{r} - \frac{kq}{r} + \frac{kq}{r} - \frac{kq}{r}$. They did not include the original charge while calculating the potential and finally wrote down $V = -\frac{kq}{r}$, $r = 2a$, $k = \frac{1}{4\pi\epsilon_0}$. As already discussed in other cases with different students, we again find that G3 did not interpret the distance in the expression of the potential correctly. In particular, several interviewed measured the distance incorrectly instead of calculating it from each charge to a point in the region of interest.

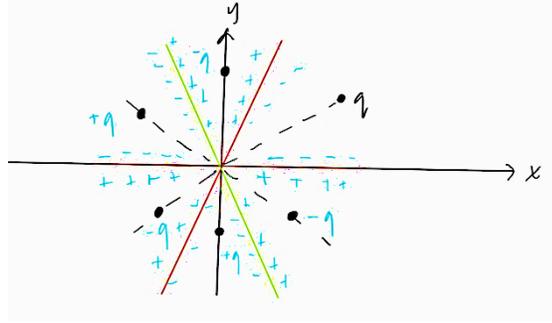

Fig. 26. A picture redrawn by G3 for Q4 to figure out the correct number, magnitude and sign of image charges

---